\documentclass{amsart}
\usepackage{graphicx}

\theoremstyle{plain}% default
\newtheorem{theorem}{Theorem}[section]

\newtheorem{cor}[theorem]{Corollary}

\theoremstyle{definition}
\newtheorem*{definition}{Definition}

\theoremstyle{remark}
\newtheorem{remark}{Remark}

\numberwithin{equation}{section}

\begin{document}

\title[Jacobi stability \ldots gravitation and cosmology]{Jacobi stability analysis of dynamical systems -- applications in gravitation and cosmology}

\author{C.~G.~B\"ohmer}
\address{Department of Mathematics and Institute of Origins, University College London, Gower Street, London, WC1E 6BT, United Kingdom}
\email{c.boehmer@ucl.ac.uk}

\author{T.~Harko}
\address{Department of Physics and Center for Theoretical and Computational Physics, The University of Hong Kong, Pok Fu Lam Road, Hong Kong}
\email{harko@hkucc.hku.hk}

\author{S.~V.~Sabau}
\address{Tokai University, Sapporo Campus, Department of Human Science and Informatics, 5-1 Minamisawa, Minamiku, Sapporo, 005-8601 Japan}
\email{sorin@tspirit.tokai-u.jp}

\def\frax #1#2{\displaystyle\frac{#1}{#2}}

\date{\today}

\begin{abstract}
The Kosambi-Cartan-Chern (KCC) theory represents a powerful mathematical method for the analysis of dynamical systems. In this approach one describes the evolution of a dynamical system in geometric terms, by considering it as a geodesic in a Finsler space. By associating a non-linear connection and a Berwald type connection to the dynamical system, five geometrical invariants are obtained, with the second invariant giving the Jacobi stability of the system. The Jacobi (in)stability is a natural generalization of the (in)stability of the geodesic flow on a differentiable manifold endowed with a metric (Riemannian or Finslerian) to the non-metric setting. In the present paper we review the basic mathematical formalism of the KCC theory, and present some specific applications of this method in general relativity, cosmology and astrophysics. In particular we investigate the Jacobi stability of the general relativistic static fluid sphere with a linear barotropic equation of state, of the vacuum in the brane world models, of a dynamical dark energy model, and of the Lane-Emden equation, respectively.  It is shown that the Jacobi stability analysis offers a powerful and simple method for constraining the physical properties of different systems, described by second order differential equations.
\end{abstract}

\date{\today}
%\pacs{04.50.+h, 04.20.Jb, 04.20.Cv, 95.35.+d}

\maketitle

\tableofcontents

\section{Introduction}

A second order differential equation can be investigated in geometric terms by using the general path-space theory of Kosambi-Cartan-Chern (KCC-theory) inspired by the geometry of a Finsler space \cite{Ko33,Ca33,Ch39}. The KCC theory is a differential geometric theory of the variational equations for the deviation of the whole trajectory to nearby ones. By associating a non-linear connection and a Berwald type connection to the differential system, five geometrical invariants are obtained. The second invariant gives the Jacobi stability of the system \cite{Sa05,Sa05a}. The KCC theory has been applied for the study of different physical, biochemical or technical systems (see \cite{Sa05,Sa05a,An93,An00,YaNa07} and references therein).

The Jacobi stability of a dynamical system can be regarded as the {\it robustness} of the system to small perturbations of the whole trajectory \cite{Sa05}. This is a very convenable way of regarding the resistance of limit cycles to small perturbation of trajectories. On the other hand, we may regard the Jacobi stability for other types of dynamical systems (like the ones in the present paper) as the resistance of a whole trajectory to the onset of chaos due to small perturbations of the whole trajectory. This interpretation is based on the generally accepted definition of chaos, namely a compact manifold $M$ on which the geodesic trajectories deviate exponentially fast. This is obviously related to the curvature of the base manifold (see section \ref{kcc}). The Jacobi (in)stability is a natural generalization of the (in)stability of the geodesic flow on a differentiable manifold endowed with a metric (Riemannian or Finslerian) to the non-metric setting. In other words, we may say that Jacobi unstable trajectories of a dynamical system behave chaotically in the sense that after a finite interval of time it would be impossible to distinguish the trajectories that were very near each other at an initial moment.

It is the purpose of the present paper to review the KCC theory, to analyze its relations with the linear Lyapunov stability analysis of dynamical systems, and to present a comparative study of these methods in the fields of gravitation and astrophysics. The Jacobi stability analysis has been extensively applied in different fields of science and technology, but presently its applications to the very important fields of gravitation, cosmology and astrophysics have been limited to the Jacobi stability analysis of the Lane-Emden equation \cite{BoHa09}, and to the analysis of the stability of the vacuum in the brane world models \cite{HaSa08}. Besides reviewing these examples, we will present in detail several new applications in gravitation and cosmology of the KCC theory, including the analysis of the general relativistic static fluid sphere, and of the interacting dark energy models. In all these cases we will also carefully consider the physical implications of our stability analysis. 

The present paper is organized as follows. In Section \ref{2} we review the basic mathematical concepts related to the analysis of the linear stability of the dynamical systems.  The KCC theory and the Jacobi stability analysis of the dynamical systems are presented in Section \ref{3}. The case of the Newtonian polytropes is considered in Section \ref{4}. The Jacobi stability analysis of the general relativistic fluid sphere is  performed in Section \ref{sec5}. The stability analysis of the vacuum in the brane world models is investigated in Section \ref{sec6}. Dynamical dark energy models are considered in Section \ref{sec7}. We discuss and conclude our review in Section \ref{sec8}. 

\section{Dynamical systems}\label{2}

\subsection{Linear stability of ODE's}

In this section we will recall some basics about nonlinear systems of ODE's. The reader familiar with these can skip this and the next subsection. Our exposition follows closely \cite{R 1995}.

Let us consider the following system of ordinary differential equations
\begin{equation}\label{3.1}
  \frax{d{\bf x}}{dt}=f({\bf x}),
\end{equation}
where ${\bf x}$ is a point in some open set $U\subset {I\!\!R}^n$ (or some differentiable manifold, like the torus, for instance), and $f:U \subset {I\!\!R}^n\to {I\!\!R}^n$ is a differentiable function. The function $f$ is also called {\it a vector field} because it assigns a vector $f(x)$ to each point in $U$. If $f$ is a linear function, then \eqref{3.1} is called {\it a linear system of ODE's}. We are however concerned with nonlinear systems of ODE's that appear in modeling real life phenomena.

Given the differential equation \eqref{3.1}, let $\varphi^t({\bf x}_0)$ be the solution ${\bf x}(t)$ with the initial condition ${\bf x}_0$ at $t=0$, i.e.
\begin{equation}
  \varphi^0({\bf x}_0)={\bf x}_0,\qquad \frax{d}{dt}\varphi^t({\bf x}_0)=f(\varphi^t({\bf x}_0)),
\end{equation}
at all $t$ for which the solution is defined. It is customary to write $\varphi(t,{\bf x}_0)$ for $\varphi^t({\bf x}_0)$. The function $\varphi^t({\bf x}_0)$ is called the {\it flow} of the differential equation, and the function $f$ defining the differential equation is called the {\it vector field which
generates the flow}.

A point ${\bf x}$ is called a {\it fixed point} (or a {\it stationary point}) for the flow $\varphi^t$ if $\varphi^t({\bf x})={\bf x}$. The point ${\bf x}$ it is also called an {\it equilibrium}, or a {\it steady-state}\index{steady-states}, or {\it singular point}. If the flow is obtained as the solution of the differential equation $\dot{{\bf x}}=f({\bf x})$, then a fixed point is a point for which $f({\bf x})=0$.

A point ${\bf x}$ is called a {\it periodic point} provided there is a time $T>0$ such that  $\varphi^T({\bf x})={\bf x}$ and $\varphi^t({\bf x})\neq {\bf x}$ for $0<t<T$. The time $T>0$ which satisfies the above conditions is called  {\it the period} of the flow.

The set $O({\bf x})=\{\varphi^t({\bf x}): t\in I_{\bf x}\}$ is called {\it the orbit} of the point ${\bf x}$, where $I_{\bf x}=(t_-,t_+)$ is the maximal interval of definition of the flow. The orbit of a periodic point ${\bf x}$ is called a  {\it periodic orbit}. Periodic orbits are also {\it closed orbits} since the set of points on the orbit is a closed curve. Furthermore, one can define the {\it positive semiorbit} and the {\it negative semiorbit} as $O^+({\bf x})=\{\varphi^t({\bf x}):t> 0\}$ and $O^-({\bf x})=\{\varphi^t({\bf x}):t< 0\}$, respectively.

The {\it positive limit set} and the {\it negative limit set} of a point ${\bf x}$ are defined as
\begin{equation}
  \Lambda^+({\bf x})=\{{\bf y}: \textrm{there exists a sequence } t_n\to \infty \textrm{ such that }\lim_{n\to \infty}\varphi(t_n,{\bf x})={\bf y}\}
\end{equation}
and
\begin{equation}
  \Lambda^-({\bf x})=\{{\bf y}: \textrm{there exists a sequence } t_n\to -\infty \textrm{ such that }\lim_{n\to \infty}\varphi(t_n,{\bf x})={\bf y}\},
\end{equation}
respectively.

Obviously, if the {\it forward orbit} $O^+({\bf x})$ is bounded, then $\Lambda^+({\bf x})$ is connected.

Let us consider an example from \cite{R 1995}:
\begin{equation}
  \dot{x}=-y+\mu x (1-x^2-y^2),\qquad \dot{y}=x+\mu y (1-x^2-y^2),
\end{equation}
for $\mu>0$. In polar coordinates the system becomes
\begin{equation}
  \dot{\theta}=1,\qquad \dot{r}=\mu r (1-r^2).
\end{equation}
The derivative of $r$ satisfies
\begin{equation}
  \dot r > 0 \ \textrm{for}\  0<r<1,
  \quad \textrm{and} \quad
  \dot r < 0 \ \textrm{for}\  1<r.
\end{equation}
Therefore, for some initial conditions $x\neq (0,0)$, $r_0\neq 0$, the trajectory has the positive limit set $\Lambda^+({\bf x})=S^1$. Thus solutions forward in time converge onto the unique periodic orbit  $r=1$.

\begin{quote}
Such an orbit with trajectories spiraling toward it is called a {\it limit cycle}. \index{limit cycle}
\end{quote}

For nonlinear systems, one can linearly approximate them near hyperbolic fixed points, by means of the Jacobian, such that a linear stability analysis holds good.

For concreteness, consider a nonlinear differential equation of the type \eqref{3.1}, and let $p$ be a fixed point, namely $\varphi^t(p)=p$ for all $t$. In order to linearize the flow $\varphi^t$ near $p$, let $A=(Df)_{|p}=(\partial f_i / \partial x_j)|_p$ be the Jacobian matrix of $f$ evaluated at $p$. Then the {\it linearized differential equation} at $p$ is $\dot{{\bf x}}=A{\bf x}$.

The fixed point $p$ is called {\it hyperbolic} if the real part of the eigenvalues of the matrix $A$ are nonzero. A hyperbolic fixed point is a {\it sink}, or said to be {\it attracting}, provided the real parts of all eigenvalues of $A$ are negative. A hyperbolic fixed point is  a {\it source}, or said to be {\it repelling}, if the real parts of all eigenvalues are positive. Finally, a  hyperbolic fixed point is called a {\it saddle} if it is neither a sink nor a source; this would be the case if there exist eigenvalues $\lambda_+$ and $\lambda_-$ such that $Re(\lambda_+)>0$ and $Re(\lambda_-)<0$. A nonhyperbolic fixed point is said to have a {\it center}.

One is often interested in the set of all points whose limit set is a fixed point. The {\it stable manifold}, or the {\it basin of attraction}, of a fixed point $p$ is the set
\begin{equation}
  W^s(p)=\{{\bf x}:\Lambda^+({\bf x})={\bf x}\},
\end{equation}
and the {\it unstable manifold} of a fixed point $p$ is the set
\begin{equation}
  W^u(p)=\{{\bf x}:\Lambda^-({\bf x})={\bf x}\}.
\end{equation}

The linearization method of Hartman says that the differential system \eqref{3.1} and the linearized system are locally topologically equivalent. Namely,

\begin{theorem}[(Hartman)~\cite{R 1995}]
Consider the differential equation \eqref{3.1} with $f$ smooth in an open subset $U$ of $\mathbb R^n$ containing the origin. If the origin ${\bf x}=0$ is a hyperbolic fixed point, then in a small neighborhood of ${\bf x}=0$, there exist stable and unstable manifolds $W^s$ and $W^u$ with the same dimensions as the stable and unstable manifolds $E^s$ and $E^u$ of the linearized system, respectively, where $W^s$ and $W^u$ are tangent to $E^s$ and $E^u$ at ${\bf x}=0$.
\end{theorem}

If the fixed point is not hyperbolic, then other methods have to be used.

\subsection{Planar Nonlinear differential equations systems}

The {\it phase-portrait} of a two-dimensional differential equation $\dot{{\bf x}}=f({\bf x})$ is a drawing of the solution curves with the direction of increasing time indicated. In an abstract sense, the phase portrait is the drawing of all solutions, but in practice it only includes the representative trajectories. The {\it phase space} is the domain of all ${\bf x}$'s considered.

\begin{definition}
Let $p$ be a fixed point of the two-dimensional system $\dot{{\bf x}}=f({\bf x})$, and denote by $\lambda_1$, $\lambda_2$ the two eigenvalues of $A := (Df)_{|p}$. The following classification of the fixed point $p$ is standard.

\begin{itemize}
\item[(1)]
 {\bf $\lambda_1$, $\lambda_2$ are real and distinct.}
 \begin{itemize}
  \item[1.1.]
   {\bf $\lambda_1\cdot \lambda_2>0$ (the eigenvalues have the same sign)}:
   $p$  is called a {\it node} or type I singularity; that is,
   every orbit tends to the origin in a definite direction as $t \to \infty$.
   \begin{itemize}
    \item[1.1.1.] {\bf $\lambda_1, \lambda_2>0$ }:
                  $p$ is an {\it unstable node}.\index{unstable node}
    \item[1.1.2.] {\bf $\lambda_1, \lambda_2<0$ }:
                  $p$ is a {\it stable node}.\index{stable node}
   \end{itemize}
  \item[1.2.]
   {\bf  $\lambda_1\cdot \lambda_2<0$ (the eigenvalues have different signs)}:
   $p$  is an {\it unstable fixed point},
   or a {\it saddle} point singularity.
 \end{itemize}
\item[(2)]
 {\bf  $\lambda_1$, $\lambda_2$ are complex,
 i.e. $\lambda_{1,2}=\alpha\pm i \beta$, $\beta \neq 0$.}
 \begin{itemize}
  \item[2.1.]
   {\bf $\alpha\neq 0$}: $p$ is a {\it spiral}, or a {\it focus}, that is, the solutions
   approach the origin as $t\to \infty$, but not from a definite direction.
   \begin{itemize}
    \item[2.1.1.] {\bf $\alpha<0$}:
                  $p$ is a {\it stable focus}.
    \item[2.1.2.] {\bf $\alpha>0$}:
                  $p$ is an {\it unstable focus}.
   \end{itemize}
  \item[2.2.] {\bf $\alpha = 0$}: $p$ is a {\it center}, \index{center}
   that means it is not stable in the usual sense, and we have to look
   at higher order derivatives.
 \end{itemize}
\item[(3)]
 {\bf $\lambda_1$, $\lambda_2$ are equal, i.e. $\lambda_1=\lambda_2=\lambda$.}
 \begin{itemize}
  \item[3.1.] If there are two linearly independent eigenvectors,  we have a {\it star singularity}, or a {\it stable singular node} (these are simple straight lines
   through the origin).
    \item[3.2.] If there is only one linearly independent eigenvector, we have an {\it improper node}, or {\it unstable degenerate node}.   %
 \end{itemize}
\end{itemize}
\end{definition}

%%%%%%%%%%%%%%%%%%%%%%%%%%%%%%%%%%%%%%%%%%%%%%%%%%%%%%%%%%%%%%
\begin{figure}[!ht]
\includegraphics[height=6cm]{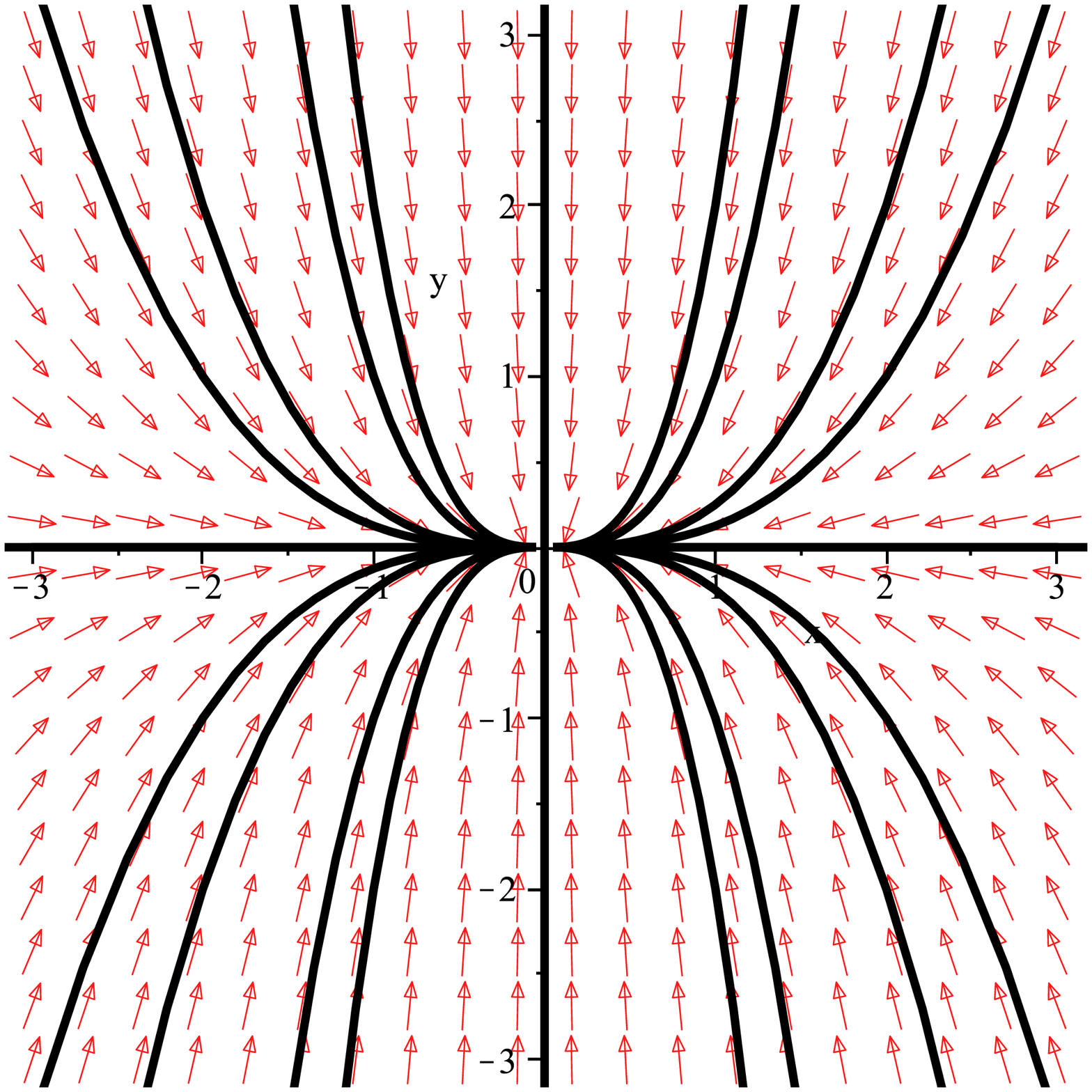}
\hfill
\includegraphics[height=6cm]{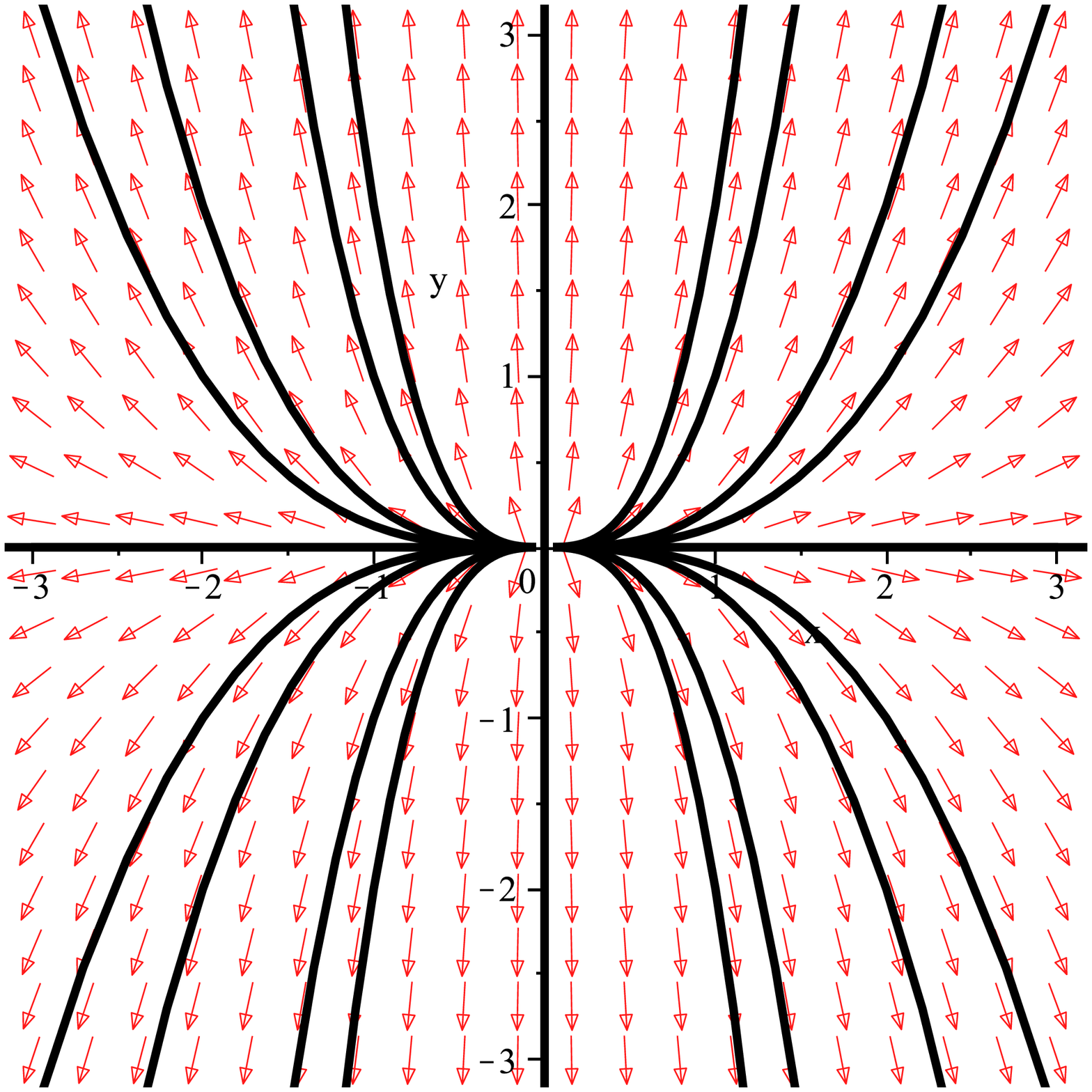}
\caption{Stable (left) versus unstable (right) nodes.}
\end{figure}
%%%%%%%%%%%%%%%%%%%%%%%%%%%%%%%%%%%%%%%%%%%%%%%%%%%%%%%%%%%%%%%
\begin{figure}[!ht]
\includegraphics[height=6cm]{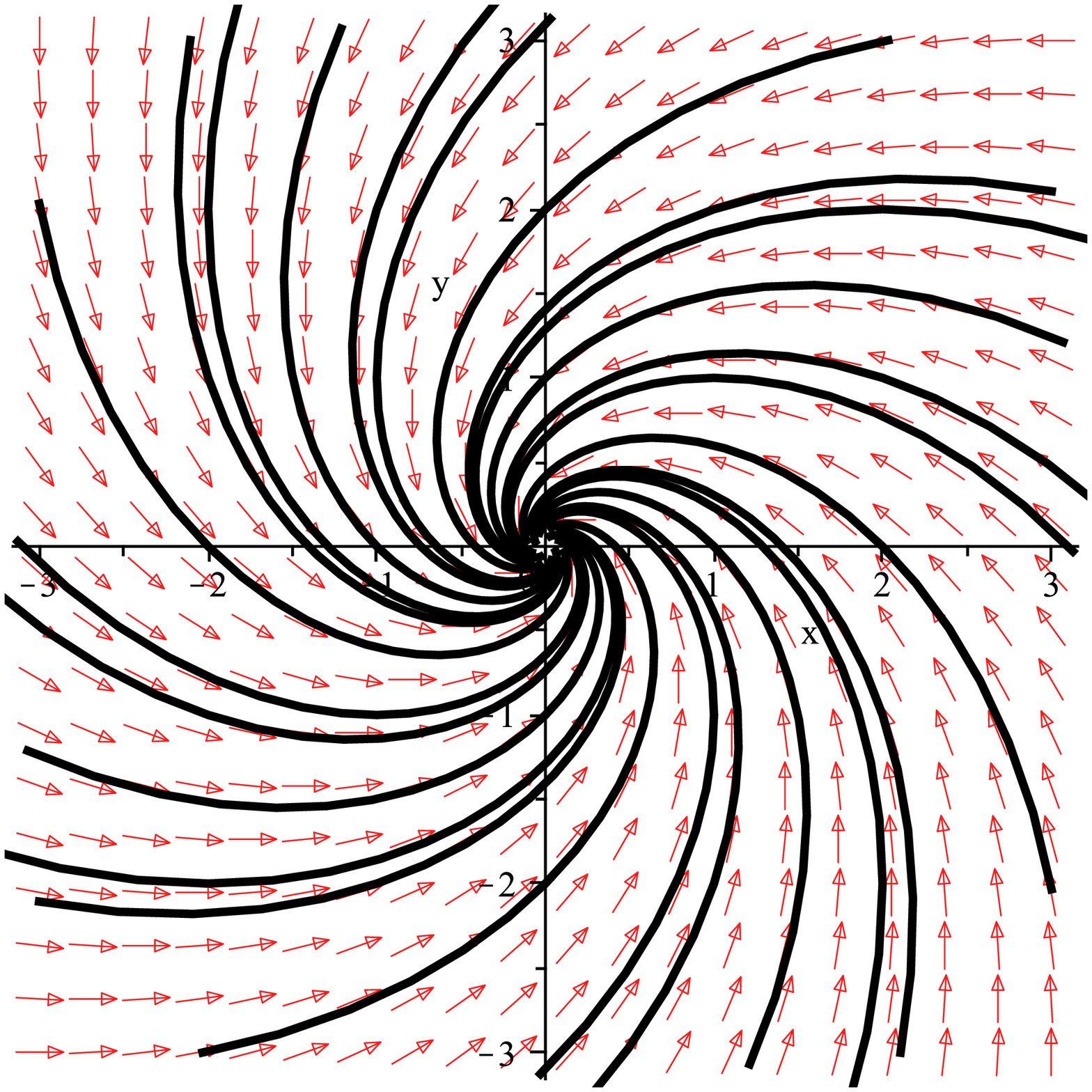}
\hfill
\includegraphics[height=6cm]{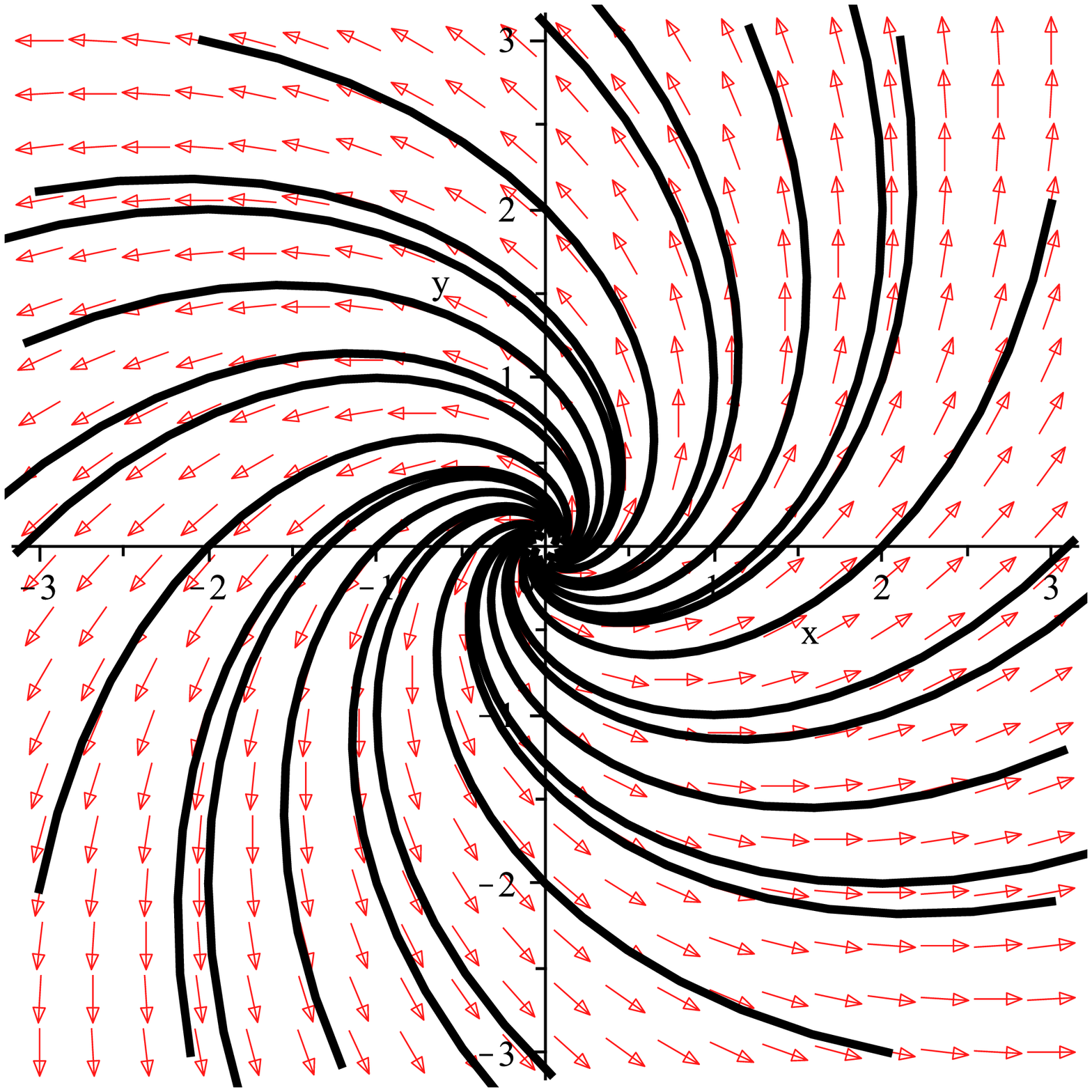}
\caption{Stable (left) versus unstable (right) spirals.}
\end{figure}
%%%%%%%%%%%%%%%%%%%%%%%%%%%%%%%%%%%%%%%%%%%%%%%%%%%%%%%%%%%%%%%
\begin{figure}[!ht]
\includegraphics[height=6cm]{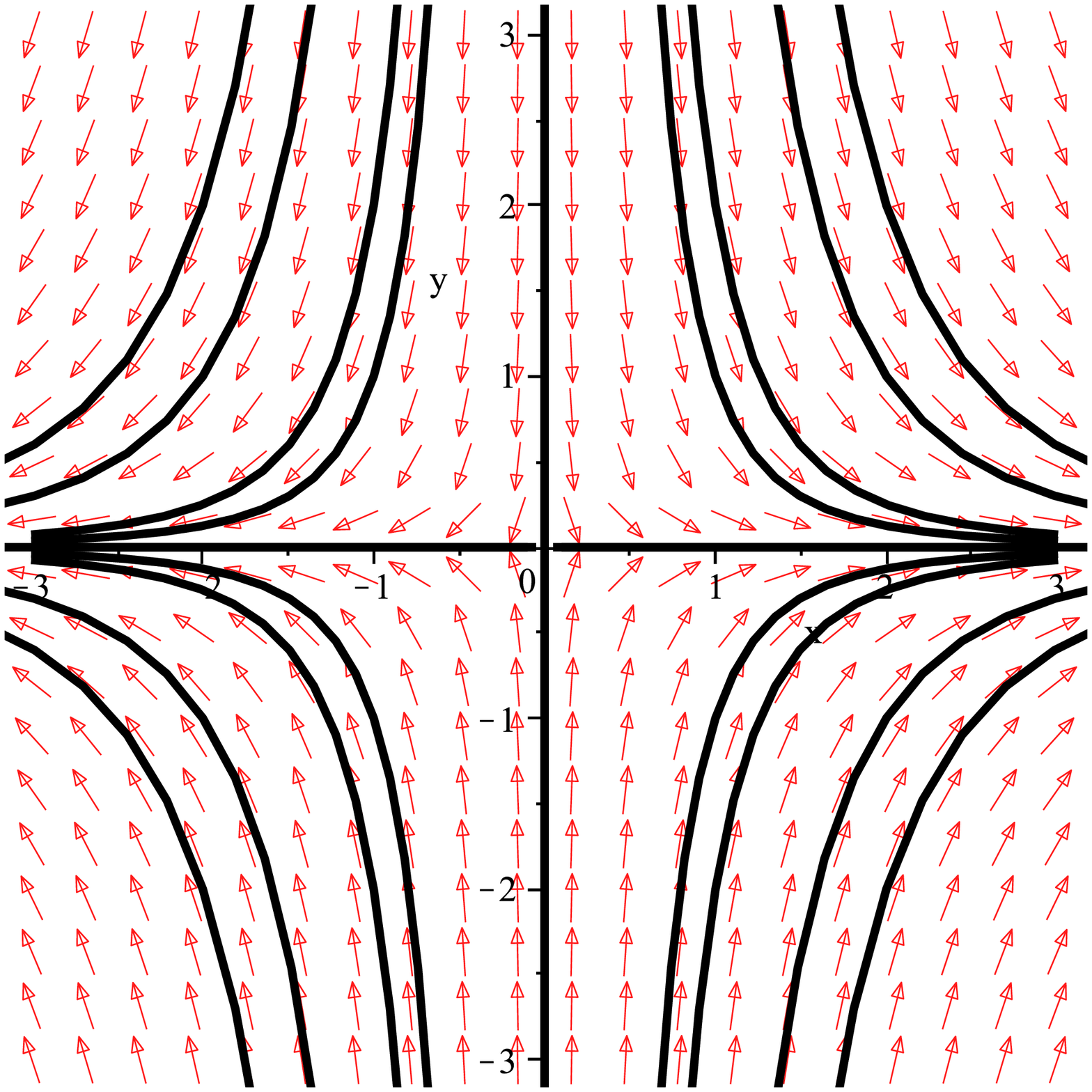}
\hfill
\includegraphics[height=6cm]{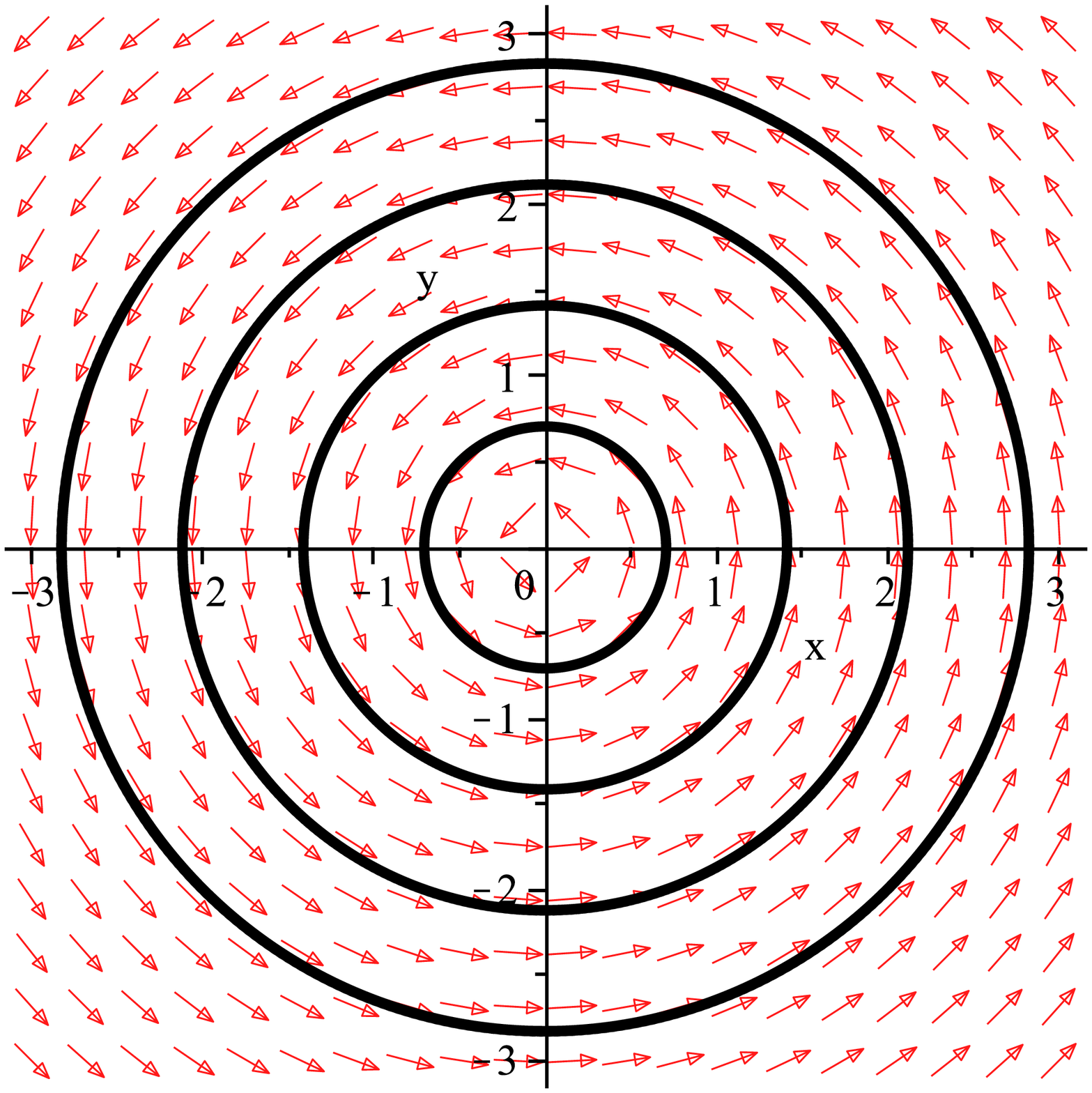}
\caption{Saddle point singularity (left) and a center (right).}
\end{figure}
%%%%%%%%%%%%%%%%%%%%%%%%%%%%%%%%%%%%%%%%%%%%%%%%%%%%%%%%%%%%%%%
\begin{figure}[!ht]
\includegraphics[height=6cm]{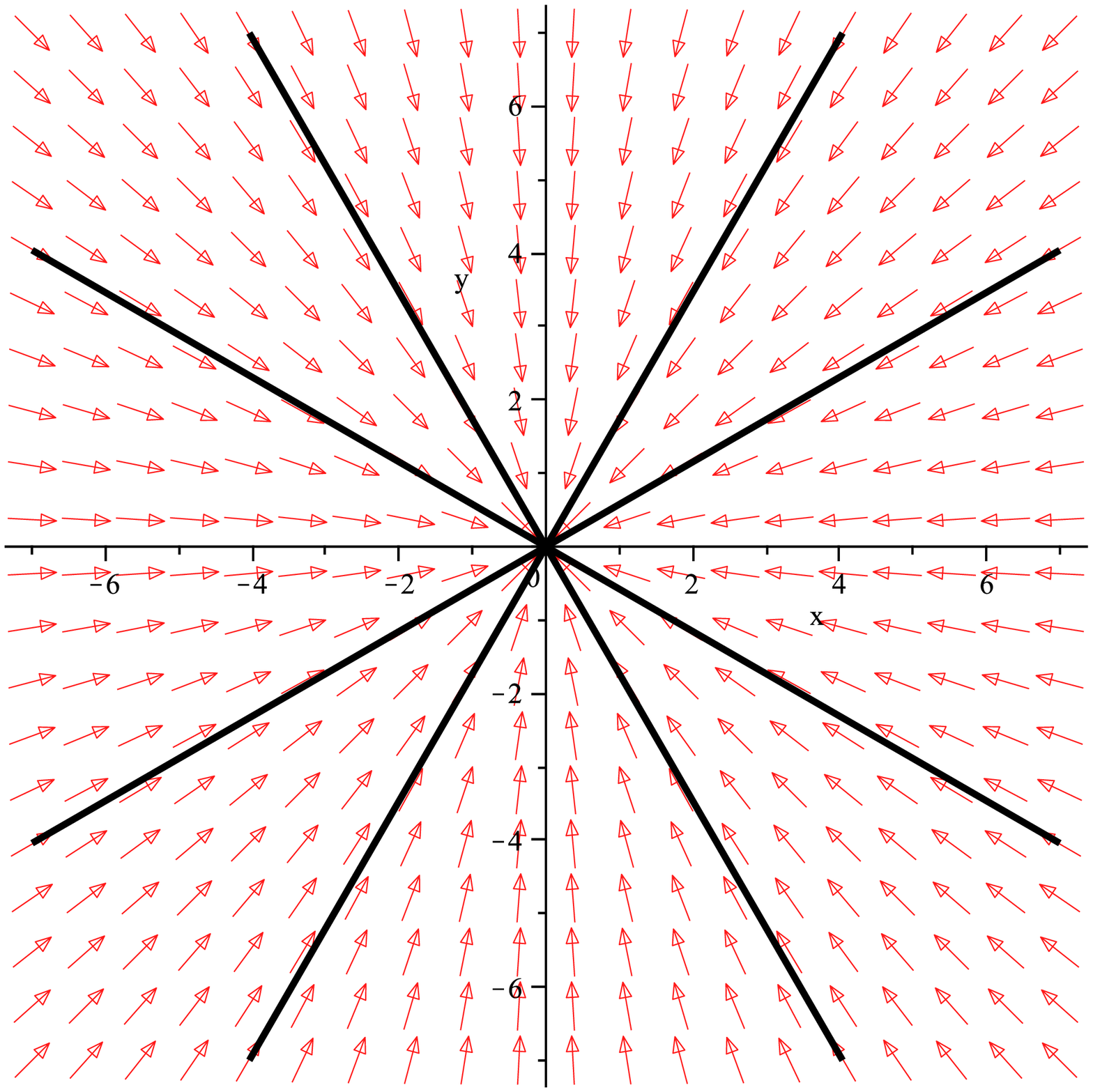}
\hfill
\includegraphics[height=6cm]{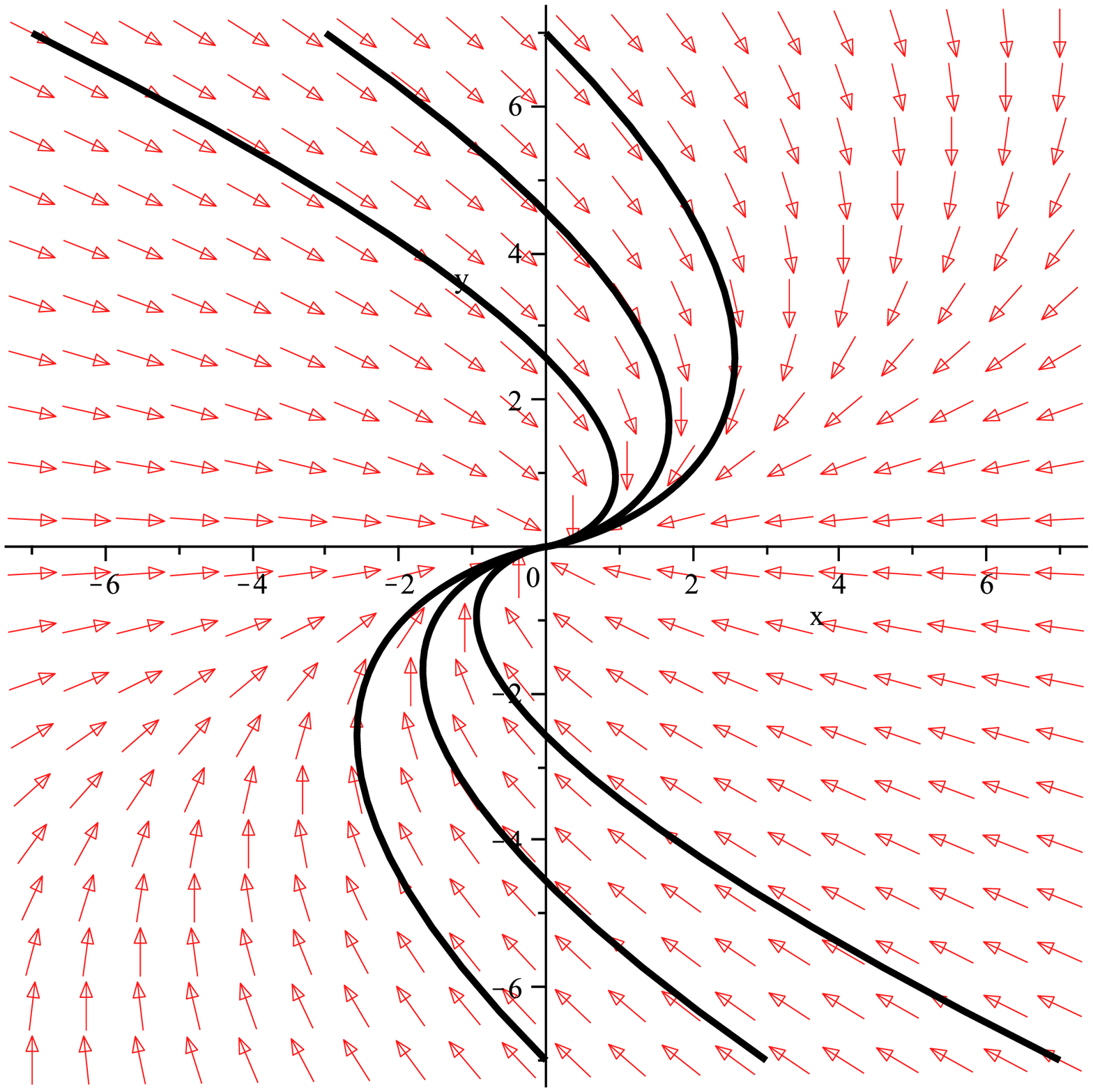}
\caption{Stable singular node (left) versus unstable degenerate node (right).}
\end{figure}
%%%%%%%%%%%%%%%%%%%%%%%%%%%%%%%%%%%%%%%%%%%%%%%%%%%%%%%%%%%%%%%%%%

\begin{remark}
As one can easily remark from the above classification, linear stability means that the fixed point $p$ is an {\it attractor} for the nearby solutions, or,  geometrically, that the trajectories are ``converging". For the sake of simplicity, let us consider here a one dimensional autonomous differential equation $\dot x=f(x)$ in the Euclidean plane $(\mathbb R, ||\cdot ||)$ with a fixed point $p$. This means that if a solution starts at this point, it remains there forever, i.e. $p(t)=p$, for all $t$.
\end{remark}

Strictly speaking, the critical point $p$ is called {\it stable} if given $\varepsilon>0$, there is a $\delta>0$ such that for all $t\geq t_0$, $||x(t)-p(t)||<\varepsilon$ whenever $t\geq t_0$, $||x(t_0)-p(t_0)||<\delta$, where  $x(t)$ is a solution of $\dot x=f(x)$. One can see that this definition is equivalent to the fact that $f'(x_0)<0$, as predicted by the general case. Indeed, let us consider a small perturbation $\xi(t)$ away from the critical point $p$, i.e.
\begin{equation}\label{linear_perturb}
  x(t)=p+\xi(t).
\end{equation}
We will analyze now the tendency of the perturbation $\xi(t)$ to grow or decay in time. By taking the derivative of \eqref{linear_perturb} it follows
\begin{equation}
  \dot \xi(t)=\dot x(t)=f(x)=f(p+\xi),
\end{equation}
and by Taylor expansion, we get
\begin{equation}
  \dot\xi(t)=f(p)+\xi f'(p)+\frac{\xi^2}{2}f''(p)+\cdots.
\end{equation}

We are interested in linear analysis, therefore the higher order terms can be neglected. Taking into account that $f(p)=0$, we obtain
\begin{equation}
  \dot\xi =\xi f'(p)
\end{equation}

We can immediately conclude from here that
\begin{enumerate}
\item if $f'(p)<0$, then the perturbation $\xi(t)$ decays exponentially, i.e. trajectories are converging, and
\item if $f'(p)>0$, then the perturbation $\xi(t)$ grows exponentially, i.e. trajectories are diverging.
\end{enumerate}

The higher dimensional case can be analogously treated leading to the classification above.

%%%%%%%%%%%%%%%%%%%%%%%%%%%%%%%%%%%%%%%%%%
%%%%%%%%%%%%%%%%%%%%%%%%%%%%%%%%%%%%%%%%%

\subsection{Limit Cycles}

Roughly speaking,  a {\it limit cycle} is an isolated periodic solution.

It is known that for a planar ODE system, the trajectories tend to a critical point, a closed orbit, or to infinity.

In order to discuss the stability of limit cycles, let us recall the notion of invariant set. A subset $S\in \mathbb R^2$ is called {\it invariant} with respect to the flow  $\varphi^t$ if, for any ${\bf x\in S}$, it follows $\varphi^t({\bf x})\in S$, for all $t\in I_{\bf x}$.
Similarly, a subset $S\in \mathbb R^2$ is called {\it positively invariant} and {\it negatively invariant} with respect to the flow  $\varphi^t$ if, for any ${\bf x\in S}$, it follows $O^+({\bf x})\in S$, and $O^-({\bf x})\in S$, respectively.

Then, a limit cycle, say $\Phi$, is called
\begin{enumerate}
\item a {\it stable limit cycle} if $\Lambda^+({\bf x})=\Phi$, for all ${\bf x}$ in some neighborhood, this obviously means that the nearby trajectories are {\it attracted} to the limit cycle;
\item an {\it unstable limit cycle} if $\Lambda^-({\bf x})=\Phi$, for all ${\bf x}$ in some neighborhood, meaning that the nearby trajectories are {\it repelled} from the limit cycle;
\item a {\it semistable limit cycle} if it is attracting on one side and repelling on the other.
\end{enumerate}

Recall here an useful tool for the study of existence of limit cycles in plane, namely the Poincar\'e-Bendixon Theorem.

\begin{theorem}[Poincar\'e-Bendixon]
Suppose $O^+{\bf x}$ is contained in a bounded region in which there are only finitely many critical points. Then $\Lambda^+(O^+)$ is one of the following
\begin{enumerate}
\item a single critical point;
\item a single closed orbit;
\item a graphic, i.e. critical points joined by heteroclinic orbits.
\end{enumerate}
\end{theorem}

In concrete problems, the following corollary is usually used

\begin{cor}
Let S be a bounded closed set containing no critical points and suppose that S is positively invariant. Then, there exists a limit cycle entirely contained in S.
\end{cor}

The eigenvalues of an appropriate matrix can be used to determine the {\it stability of periodic orbits, or limit cycles}. This is the so-called {\it Floquet theory}.

We are interested in the study of dynamics of a flow in a neighborhood of a periodic orbit. One method to do this is to follow the nearby trajectories as they make one circuit around the periodic orbit.

Let $\varphi$ be a periodic orbit of period $T$ and $p\in \varphi$, i.e.~$\varphi^T(p)=p$. Then for some $k\in {\mathbb N}$, the $k$-th coordinate function of the vector field must not vanish at $p$, i.e. there exists a $k$ such that $f_k(p)\neq 0$. We consider the hyperplane $\Sigma=\{{\bf x} | \, {\bf x}_k=p_k\}$, called a {\it cross section} or {\it transversal} at $p$.

For ${\bf x}\in \Sigma$ near $p$ the flow $\varphi^t({\bf x})$ returns to $\Sigma$ in time $\tau({\bf x})$, which is roughly equal to $T$ (this follows from the Implicit Function Theorem).

\begin{figure}[!ht]
\centering
\includegraphics[height=6 cm]{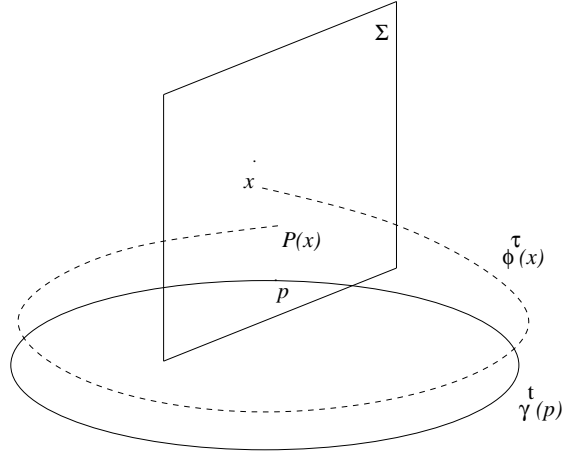}
%ymsriSapmap.eps
\caption{The Poincar\' e map.}
\label{poincaremap}
\end{figure}

Let $V\subset \Sigma$ be an open set on which $\tau({\bf x})$ is a differentiable function. The {\it first return map} or {\it Poincar\'e map} \index{Poincar\'e map} $\mathcal P:\Sigma \to \Sigma$ is defined to be $\mathcal P({\bf x})=\varphi^{\tau({\bf x})}({\bf x})$ for ${\bf x}\in V$, see Fig.~\ref{poincaremap}.

It can be seen that the Poincar\'e map is differentiable. If $\varphi$ is a periodic orbit of period $T$ with $p\in \varphi$ then it can be shown that the eigenvalues of $(D\varphi^T)_{|{\bf x}=p}$ are $1,\lambda_1,...,\lambda_{n-1}$ (the eigenvector of 1 is $f(p)$). The $n-1$ eigenvalues $\lambda_1, ...,  \lambda_{n-1}$ are called {\it characteristic multipliers} of the periodic orbit $\varphi$. It is remarkable that these depend only on $\varphi$ but not on the point $p$.

The central result of Floquet theory says that the eigenvalues of the Poincar\'e map's Jacobian matrix at $p$ (i.e. $(D \mathcal P)_{|x=p}$) coincide with the characteristic multipliers of the limit cycle. Based on this result, the stability of limit cycles has been formulated in terms of their characteristic multipliers.

A limit cycle $\varphi$ is {\it hyperbolic} provided $|\lambda_j|\neq 1$ for all characteristic multipliers ($1\leq j\leq n-1$). It is {\it a hyperbolic stable limit cycle}, a {\it periodic attractor}, or a {\it periodic sink}, if all $|\lambda_j|<1 $. It is {\it a hyperbolic unstable stable limit cycle}, a {\it periodic repeller}, or a {\it periodic source}, if all $|\lambda_j|>1 $. A hyperbolic periodic orbit which is neither a source or a sink is called a {\it saddle limit cycle}.

\begin{remark}[The planar systems case]

For a two dimensional differential equation the Poincar\'e map \index{Poincar\'e map} can sometimes be explicitly calculated. Indeed, let us consider the differential equation
\begin{equation}
  \dot{x}=X(x,y) \qquad \dot{y}=Y(x,y)
\end{equation}
and
\begin{equation}
  V(x,y) = \left( \begin{array}{l}
    X(x,y)\\
    Y(x,y)
  \end{array} \right).
\end{equation}

Let us  consider the transversal $\Sigma$ and let $\Sigma '\subset \Sigma$ be the open set where Poincar\'e map is defined, $\mathcal P : \Sigma' \to \Sigma$. Since we are in a plane, $\Sigma$ is a curve which can be reparametrized by $\gamma:I\to \Sigma$ with $\gamma(I')=\Sigma '$, $I'\subset I$ and $|\gamma(s)|=1$. As in the general case, let $\tau(q)$ be the return time for $q\in \Sigma'$, so $\mathcal P(q)=\varphi^{\tau(q)}(q)$.

If we assume $\mathcal P(q_0)=q_0$ and $\gamma(s_0)=q_0$, then
\begin{equation}
  (\mathcal P \circ \gamma)'(s_0)
  =\exp[\int_0^{\tau(q_0)}( div \ V)\circ \varphi^t(q_0)dt]
\end{equation}
(see \cite{R 1995} for details).

In this case, one can define the {\it characteristic multiplier M} to be
\begin{equation}
  M:=\frac{d\mathcal P}{dq}_{|q^*},
\end{equation}
where $q^*$ is a fixed point of the Poincar\'e map $\mathcal P$ corresponding to a limit cycle, say $\Phi$. If $|M|<1$, then $\Phi$ is a hyperbolic stable limit cycle, if $|M|>1$, then $\Phi$ is an  unstable limit cycle, and if $|M|=1, \frac{d^2\mathcal P}{dq^2}\neq 0$, then $\Phi$ is a semistable limit cycle.
\end{remark}

 Let us also recall here that, for an ODE system depending on a parameter $k$, i.e.,
 \begin{equation}
   \frac{d{\bf x}}{dt}=f({\bf x},k),
 \end{equation}
 a value, say $k_0$ is called {\it bifurcation value} if at the point $({\bf x},k)$ the stability of the solution changes. For planar systems, there are several types of bifurcation at nonhyperbolic critical points. A bifurcation concerning the stability of a limit cycle is called a {\it Hopf bifurcation}.  This is the kind of bifurcation we are most interested in. Basically, there are two types of Hopf bifurcation:
 \begin{enumerate}
 \item the {\it supercritical Hopf bifurcation} when stable limit cycles are created about an unstable critical point, and
 \item the {\it subcritical Hopf bifurcation} when an unstable limit cycle is created about a stable critical point.
 \end{enumerate}

 Finally, we mention that the problem of determining the maximum number of limit cycles for planar differential systems is still an open problem. This is the second part of the $23^{\rm rd}$ Hilbert problem and only little progress has been made to date.

\section{Kosambi-Cartan-Chern (KCC) theory and Jacobi stability}\label{3}

\subsection{General Theory}
\label{kcc}

We recall the basics of KCC-theory to be used in the sequel. Our exposition follows \cite{Sa05}.

Let $\mathcal{M}$ be a real, smooth $n$-dimensional manifold and let $T\mathcal{M}$ be its tangent bundle (in the most cases in the present exposition we will consider the case $\mathcal{M}=\mathbb R^n$). Let $u=(x,y)$ be a point in $T\mathcal{M}$, where $x=\left(
x^{1},x^{2},...,x^{n}\right)$, and $y =\left(y^{1},y^{2},...,y^{n}\right) $.

A time independent coordinate change on  $T\mathcal{M}$ is given by
\begin{equation} \label{ct}
\begin{split}
& \tilde{x}^{i}=\tilde{x}^{i}\left( x^{1},x^{2},...,x^{n}\right) \\
&  \tilde{y}^{i}=\frac{\partial \tilde{x}^{i}}{\partial x^j}y^j,
\end{split}\qquad\qquad i\in \left\{1 ,2,...,n\right\}.
\end{equation}

Let us also recall that the equations of motion of a Lagrangian mechanical system $\left( M,L,F_{i}\right)$ can be described by
 the famous {\it Euler-Lagrange equations}
\begin{equation}\label{EL}
\begin{split}
  & \frac{d}{dt}\frac{\partial L}{\partial y^{i}}-
  \frac{\partial L}{\partial x^{i}}=F_{i}\\
  & y^i=\frac{dx^i}{dt},
  \end{split}\qquad \qquad i=1,2,...,n,
\end{equation}
where $L$ is the Lagrangian of $\mathcal{M}$, and
$F_{i}$ are the external forces \cite{MiFr05,MHSS}.

For a regular Lagrangian $L$, the Euler-Lagrange equations given
by Eq. (\ref{EL}) are equivalent to a system of second-order differential
equations
\begin{equation}
  \frac{d^{2}x^{i}}{dt^{2}}+2G^{i}\left( x,y\right)=0,\qquad
  i\in \left\{ 1,2,...,n\right\},
  \label{EM}
\end{equation}
where $G^{i}\left( x,y\right) $ are smooth functions defined in a local system of coordinates on
$T\mathcal{M}$.

 The system given by
Eq.~(\ref{EM}) is equivalent to a vector field (semispray) $S$ on $T\mathcal{M}$, where
\begin{equation}
S=y^{i}\frac{\partial }{\partial x^{i}}-2G^{i}\left( x^{j},y^{j},t\right)
\frac{\partial }{\partial y^{i}},
\end{equation}
which determines a non-linear connection $N$ on $T\mathcal{M}$ with local coefficients $N_{j}^{i}$ defined by \cite{MHSS}
\begin{equation}
N_{j}^{i}=\frac{\partial G^{i}}{\partial y^{j}}.
\end{equation}
%Added by Sabau

\bigskip

More generally, one can start from an arbitrary system of
second-order differential equations on the form (\ref{EM}), where
no \textit{a priori} given Lagrangean function is assumed, and
study the behavior of its trajectories by analogy with the
trajectories of the Euler-Lagrange equations.

For a non-singular coordinate transformations given by Eq. (\ref{ct}), we
define the {\it KCC-covariant differential} of a vector field $\xi=\xi ^{i}(x)\frac{\partial}{\partial x^i}$ in an
open subset $\Omega \subseteq R^{n}\times R^{n}$ as \cite
{An93,An00,Sa05,Sa05a}
\begin{equation}
\frac{D\xi ^{i}}{dt}=\frac{d\xi ^{i}}{dt}+N_{j}^{i}\xi ^{j}.  \label{KCC}
\end{equation}

For $\xi ^{i}=y^{i}$ we obtain
\begin{equation}
\frac{Dy^{i}}{dt}=N_{j}^{j}y^{j}-2G^{i}=-\epsilon ^{i}.
\end{equation}
The contravariant vector field $\epsilon ^{i}$ on $\Omega $ is called
the {\it first KCC invariant}. We point out that the term 'invariant' refers to the fact that   $\epsilon ^{i}$ has geometrical character, namely, with respect to a coordinate transformation \eqref{ct}, we have
\begin{equation}
\tilde{\varepsilon}^i=\frac{\partial \tilde x^i}{\partial x^j}\varepsilon^j.
\end{equation}

Let us now vary the trajectories $x^{i}(t)$ of the system (\ref{EM}) into
nearby ones according to
\begin{equation}
\tilde{x}^{i}\left( t\right) =x^{i}(t)+\eta \xi ^{i}(t),  \label{var}
\end{equation}
where $\left| \eta \right| $ is a small parameter and $\xi ^{i}(t)$ are the
components of some contravariant vector field defined along the path $%
x^{i}(t)$. Substituting Eqs. (\ref{var}) into Eqs. (\ref{EM}) and taking the
limit $\eta \rightarrow 0$ we obtain the variational equations \cite
{An93,An00,Sa05,Sa05a}
\begin{equation}
\frac{d^{2}\xi ^{i}}{dt^{2}}+2N_{j}^{i}\frac{d\xi ^{j}}{dt}+2\frac{\partial
G^{i}}{\partial x^{j}}\xi ^{j}=0.  \label{def}
\end{equation}

By using the KCC-covariant differential we can write Eq. (\ref{def}) in the
covariant form
\begin{equation}
\frac{D^{2}\xi ^{i}}{dt^{2}}=P_{j}^{i}\xi ^{j},  \label{JE}
\end{equation}
where we have denoted
\begin{equation}
P_{j}^{i}=-2\frac{\partial G^{i}}{\partial x^{j}}-2G^{l}G_{jl}^{i}+ y^{l}%
\frac{\partial N_{j}^{i}}{\partial x^{l}}+N_{l}^{i}N_{j}^{l},
\end{equation}
and $G_{jl}^{i}\equiv \partial N_{j}^{i}/\partial y^{l}$ is called the
Berwald connection \cite{An93,MHSS,Sa05,Sa05a}. Eq. (\ref{JE}) is called the
{\it Jacobi equations}, or {\it the variation equations} attached to the system of second order differential equations, and $P_{j}^{i}$ is called {\it the second KCC-invariant} or {\it the
deviation curvature tensor}. When the system (\ref{EM}) describes the
geodesic equations in either Riemann or Finsler geometry, Eq. (\ref{JE}) is
the usual Jacobi equation.

The third, fourth and fifth invariants of the system (\ref{EM}) are given by
\cite{An00}
\begin{equation}
P_{jk}^{i}\equiv \frac{1}{3}\left( \frac{\partial P_{j}^{i}}{\partial y^{k}}-%
\frac{\partial P_{k}^{i}}{\partial y^{j}}\right),\qquad P_{jkl}^{i}\equiv \frac{%
\partial P_{jk}^{i}}{\partial y^{l}},\qquad D_{jkl}^{i}\equiv \frac{\partial
G_{jk}^{i}}{\partial y^{l}}.
\end{equation}

The third invariant is interpreted as a torsion tensor, while the fourth and
fifth invariants are the Riemann-Christoffel curvature tensor, and the
Douglas tensor, respectively \cite{An00}. These tensors
always exist and they describe the geometrical properties of a system of
second-order differential equations \eqref{EM} (\cite{An00,Sa05,Sa05a}).

\begin{definition}
The trajectories of (\ref{EM}) are Jacobi stable if and only if the real parts of the eigenvalues of the deviation tensor $P_{j}^{i}$ are strictly negative everywhere, and Jacobi unstable, otherwise.
\end{definition}

Let us discuss the geometrical meaning of this definition in the context of an Euclidean, Riemannian or Finslerian structure.

Since in many physical applications we are interested in the behavior of the
trajectories of the system (\ref{EM}) in a vicinity of a point $x^{i}\left(
t_{0}\right) $, where for simplicity one can take $t_{0}=0$, we will
consider the trajectories $x^{i}=x^{i}(t)$ as curves in the Euclidean space $%
\left( R^{n},\left\langle .,.\right\rangle \right) $, where $\left\langle
.,.\right\rangle $ is the canonical inner product of $R^{n}$. As for the
deviation vector $\xi $ we assume that it satisfies the initial conditions $%
\xi \left( 0\right) =O$ and $\dot{\xi}\left( 0\right) =W\neq O$, where $O\in
R^{n}$ is the null vector \cite{Sa05,Sa05a}.

For any two vectors $X,Y\in R^{n}$ we define an adapted inner product $%
\left\langle \left\langle .,.\right\rangle \right\rangle $ to the deviation
tensor $\xi $ by $\left\langle \left\langle X,Y\right\rangle \right\rangle
:=1/\left\langle W,W\right\rangle \cdot \left\langle X,Y\right\rangle $. We
also have $\left| \left| W\right| \right| ^{2}:=\left\langle \left\langle
W,W\right\rangle \right\rangle =1$.

Thus, the focusing tendency of the trajectories around $t_{0}=0$ can be
described as follows: if $\left| \left| \xi \left( t\right) \right| \right|
<t^{2}$, $t\approx 0^{+}$, the trajectories are bunching together, while if $%
\left| \left| \xi \left( t\right) \right| \right| >t^{2}$, $t\approx 0^{+}$,
the trajectories are dispersing \cite{Sa05,Sa05a}. In terms of the deviation
curvature tensor the focusing tendency of the trajectories can be described
as follows: The trajectories of the system of equations (\ref{EM}) are
bunching together for $t\approx 0^{+}$ if and only if the real part of the
eigenvalues of $P_{j}^{i}\left( 0\right) $ are strictly negative, and they
are dispersing if and only if the real part of the eigenvalues of $%
P_{j}^{i}\left( 0\right) $ are strict positive \cite{Sa05,Sa05a}.

\begin{figure}[!ht]
\centering
\includegraphics[height=6cm,width=12cm]{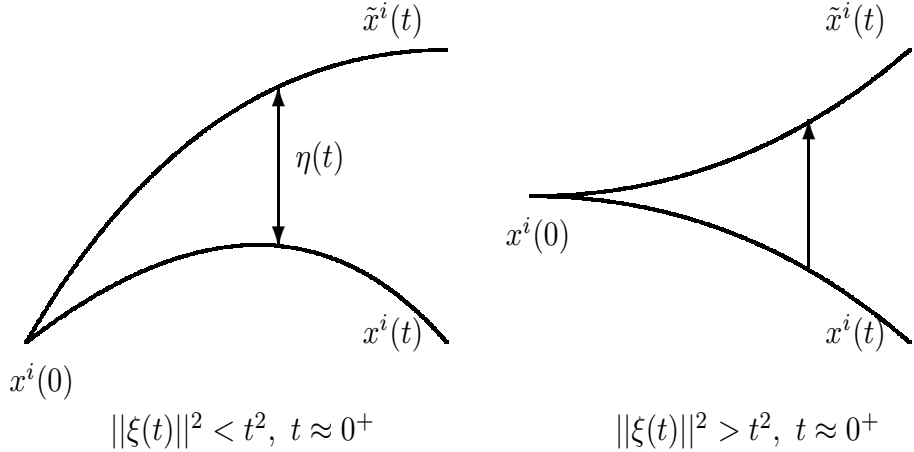}
\caption{Behavior of trajectories near zero.}
\label{pict1}
\end{figure}

The focussing behavior of the trajectories near the origin is represented in
Fig.~\ref{pict1}.

In the more general case of a typical SODE of the form \ref{EM} (not necessarily associated to any metric), let us remark that Jacobi instability via equation  \eqref{JE} means exponential growth of the deviation vector field $\xi$, while Jacobi stability means that $\xi$ have an oscillatory variation expressed by a linear combination of trigonometric functions $\sin$ and $\cos$.

%%%%%%%%%%%%%%%%%%%%%%%%%%%%%%%%%%%%%%%%%%%%%%%
%%%%%%%%%%%%%%%%%%%%%%%%%%%%%%%%%%%%%%%%%%%%%%%%

\subsection{The relation between linear stability and Jacobi stability}

It would be interesting to correlate linear stability with Jacobi stability. In other words, to compare the signs of the eigenvalues of the Jacobian  matrix $J$ at a fixed point with the signs of the eigenvalues of the deviation curvature tensor $P_i^j$ evaluated at the same point. Even though this should be possible in the general case, we give here only a discussion for the 2-dimensional case.

Let us consider the following system of ODE:
\begin{equation}\label{3.18}
  \frax{du}{dt}=f(u,v)\qquad \frax{dv}{dt}=g(u,v)
\end{equation}
such that the point $(0,0)$ is a fixed point, i.e. $f(0,0)=g(0,0)=0$. In
general, even though the fixed point is $(u_0,v_0)$, by a change of
variables $\bar{u}=u-u_0$, $\bar{v}=v-v_0$, one can obtain in most of the cases
$(0,0)$ as a fixed point. We denote by $J$ the Jacobian matrix of
\eqref{3.18}, i.e.
\begin{equation}
  J(u,v) =
  \left( \begin{array}{ll}
    f_u & f_v \\
    g_u & g_v
  \end{array} \right)
\end{equation}
where the subscripts indicate partial derivatives with respect to $u$ and $v$.

The characteristic equation reads
\begin{equation}
\lambda^2-\textrm{tr} A+\textrm{det} A=0,
\end{equation}
where
 tr $A$ and det $A$ are the trace and the determinant of the  matrix $A:=J\Big|_{(0,0)}$, respectively.

The signs of the trace, determinant of $A$, and of the discriminant $\Delta=(f_u-g_v)^2+4f_vg_u=(tr A)^2-4 \, \det A$ give the linear stability of the fixed point $(0,0)$ as described in the previous section.

Let us show how is possible to apply the KCC theory to the ODE system \eqref{3.18}. By elimination of one of the variables, we can transform \eqref{3.18} into a SODE of the form  \eqref{EM} and compute the deviation curvature. It does not matter which of the two variables we eliminate, because the resulting differential equations are topologically conjugate (\cite{R 1995}).

Let us relabel $v$ as $x$, and $g(u,v)$ as $y$, and let us assume that $g_u|_{(0,0)}\neq 0$. The variable to be eliminated, $u$ in this case, should be chosen such that this condition holds. Since $(u,v)=(0,0)$ is a fixed point, the Theorem of Implicit functions implies that the equation $g(u,x)-y=0$ has a solution $u=u(x,y)$ in the vicinity of $(x,y)=(0,0)$. Now, $\ddot x = \dot g = g_u \, f + g_v \, y$.  Hence we obtain an autonomous one-dimensional case of the SODE \eqref{EM}, namely $\ddot x^1 + g^1 = 0$, where
\begin{equation}
  g^1(x,y)=-g_u(u(x,y),x) \, f(u(x,y),x) - g_v(u(x,y),x) \, y.
\end{equation}

Evaluating now the curvature tensor $P^1_1$ from \eqref{JE} at the fixed
point $(0,0)$, we obtain after some computation:

\begin{theorem}
Let us consider the ODE \eqref{3.18} with the fixed point p=(0,0) such that $g_u|_{(0,0)}\neq 0$.

Then,
\begin{equation}
  4 \, P_1^1|_{(0,0)}
  =-4g^1_{,1}|_{(0,0)}+(g^1_{;1})^2|_{(0,0)}
  = \Delta
  := (tr A)^2 - 4 \, det A .
\end{equation}
In this case, the trajectory $v=v(t)$ is Jacobi stable if and only if  $\Delta<0$.
\end{theorem}

It can be checked that the second equality would not change if one had used the first equation of  \eqref{3.18} to eliminate the variable $v$ instead by labeling $u$ as $x$, provided $f_v|_{(0,0)}\neq 0$.

In other words, we can conclude that in the case $g_u|_{(0,0)}\neq 0$, $f_v|_{(0,0)}\neq 0$ both trajectories $u=u(t)$ and $v=v(t)$ are Jacobi stable if and only if  $\Delta<0$.

Now, straightforward calculations give the following

\begin{cor}
Let us consider the ODE \eqref{3.18} with the fixed point p=(0,0) such that $g_u|_{(0,0)}\neq 0$.

Then, the Jacobian J estimated at p has complex eigenvalues  if and only if p is a Jacobi stable point.
\end{cor}

\begin{figure}[!ht]
\centering
\includegraphics[height=6cm,width=7cm]{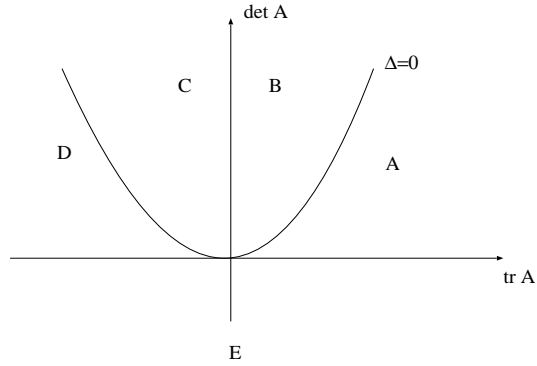} %\centering
\caption{Linear stability.}
\label{pict1_1}
\end{figure}

One can represent the linear stability as in the Fig.~\ref{pict1_1},
and by combining with the definition of Jacobi stability, in the conditions of Theorem above, we obtain the following relations
\begin{enumerate}
\item {\bf Region A.}
\begin{equation*}
%\begin{split}
\begin{matrix}
\Delta>0 &  \qquad &\textrm{Jacobi unstable}\\
S=\textrm{tr}A>0 & \qquad  & \textrm{Unstable node}\\
 P=\textrm{det}A>0
\end{matrix}
\end{equation*}
%---------------------------------------------------------------------------------------------------
\item {\bf Region B.}
\begin{equation*}
%\begin{split}
\begin{matrix}
\Delta<0 &  \qquad &\textrm{Jacobi stable}\\
S=\textrm{tr}A>0 & \qquad  & \textrm{Unstable focus}\\
 P=\textrm{det}A>0
\end{matrix}
\end{equation*}
%---------------------------------------------------------------------------------------------------
\item {\bf Region C.}
\begin{equation*}
%\begin{split}
\begin{matrix}
\Delta<0 &  \qquad &\textrm{Jacobi stable}\\
S=\textrm{tr}A<0 & \qquad  & \textrm{Stable focus}\\
 P=\textrm{det}A>0
\end{matrix}
\end{equation*}
%---------------------------------------------------------------------------------------------------
\item {\bf Region D.}
\begin{equation*}
%\begin{split}
\begin{matrix}
\Delta>0 &  \qquad &\textrm{Jacobi unstable}\\
S=\textrm{tr}A<0 & \qquad  & \textrm{Stable node}\\
 P=\textrm{det}A>0
\end{matrix}
\end{equation*}
%---------------------------------------------------------------------------------------------------
\item {\bf Region E.}
\begin{equation*}
\begin{matrix}
\Delta>0 &  \qquad &\textrm{Jacobi unstable}\\
P=\textrm{det}A<0 & \qquad  & \textrm{Saddle point}
\end{matrix}
\end{equation*}
\end{enumerate}

Let us point out that the stability in Jacobi sense refers to a linear stability type of the trajectories in {\bf the curved space} endowed with a nonlinear connection and a curvature tensor, as described above. Here the role of usual partial derivative is played by the covariant derivative along the flow. This obviously leads to difference in the meaning of linear stability and Jacobi stability.

%%%%%%%%%%%%%%%%%%%%%%%%%%%%%%%%%%%%%%%%%%%%%%%%%%%%%
%%%%%%%%%%%%%%%%%%%%%%%%%%%%%%%%

\subsection{Example: The Brusselator}

We present a first example of a dynamical system
that has a limit cycle behavior, but which is not Jacobi
stable in its entirety. In other words,
the Brusselator is not robust in
its entirety.

The Brusselator is a simple biological system
proposed by Prigogine and %%@
Lefever in 1968 \cite{Pri68} (the name Brusselator came from Brussels, the city where they
lived). The chemical reactions are:
\begin{equation}\label{B2}
\begin{array}{rll}
A    & {\stackrel{k_1}{\longrightarrow }} & X, \\
B+X  & {\stackrel{k_2}{\longrightarrow}}  & Y+D,\\
2X+Y & {\stackrel{k_3}{\longrightarrow }} & 3X,\\
X    & {\stackrel{k_4}{\longrightarrow}}  & E
\end{array}
\end{equation}
where the $k$'s are rate constants, and the reactant concentrations of $A$
and $B$ are kept constant. Then,
the law of mass action \index{law of mass action}leads to the
following system of differential equations:
\begin{equation}
\begin{array}{l}
\frax{du}{dt}=1-(b+1)u+au^2v=:f(u,v)\\
\\
\frax{dv}{dt}=bu-au^2v=:g(u,v),
\end{array}
\end{equation}
where $u$ and $v$ correspond to the concentrations of $X$ and $Y$,
respectively, and $a$, and $b$ are positive constants.

\subsubsection{ Stability of fixed points}
From the equations \eqref{B2} it follows that there is a unique fixed point
$S(u_0=1,v_0=\frax{b}{a})$.

Simple computations show that
the Jacobian evaluated at the fixed point $S$ is given by
\begin{align*}
A
&= J\vert_{(u_0,v_0)}\\
&=
\left( \begin{array}{ll}
-(b+1)+2auv & au^2 \\
b-2auv      & -au^2
\end{array} \right) \vert_{u_0=1,v_0=\frac{b}{a}} \\
&=
\left( \begin{array}{ll}
 b-1 & a \\
 -b  & -a
\end{array} \right),
% \leqno(4.3)
\end{align*}
and hence we have
\begin{equation}
\textrm{tr} A=b-1-a, \ \ \textrm{det} A=a,\ \ \Delta=(b-1-a)^2-4a.
\end{equation}

\begin{figure}[!ht]
\centering
\includegraphics[height=6cm,width=7cm]
{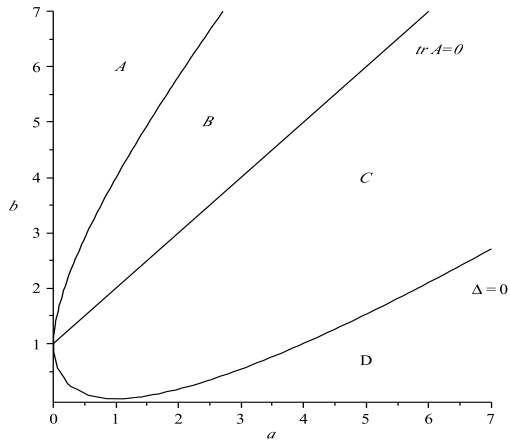} %\centering
\caption{ The parametric regions indicating the linear stability domains of the fixed point $S$ of
the Brusselator.}
\label{pict2}
\end{figure}

A graphical representation of these functions of $a$ and $b$ is given in Fig.~\ref{pict2}. Taking into account that $\textrm{det} A=a>0$, a similar analysis with the one done in the previous section gives the following table.

\begin{table}
\begin{center}
\begin{tabular}{|c|ccccccccc|}
\hline
{Region} &  & A & \big| & B & \big| & C & \big| & D &  \\
\hline
   Linear  &  &Unstable  & \big| & Unstable & \big| & Stable & \big| & Stable &\\
   stability  & &  node& \big| & focus & \big| & focus   & \big| & node &\\
   \hline
   Jacobi  &  &Jacobi  & \big| & Jacobi & \big| & Jacobi & \big| & Jacobi &\\
   stability  & &  unstable & \big| & stable & \big| & stable   & \big| & unstable &\\
    \hline
\end{tabular}
\end{center}
\caption{Linear stability versus Jacobi stability for the Brusselator system.}%\label{table1}
\end{table}

We would like to illustrate this discussion with a concrete motivation of the apparent discrepancies between linear and Jacobi stability of fixed points. For instance, if the fixed points $S$ belongs to Region D, then it must be a stable node, i.e. trajectories are attracted to $S$ in the phase plane, while they are divergent in the Jacobi stability space. A quick look at the previous section show the main differences between the variation of the deviation vector field $\xi^i(t)$ given by the formulas \eqref{def} and \eqref{JE}, respectively.  Obviously the local behavior of the solutions of these two equations are completely different. See figures

%%%%%%%%%%%%%%%%%%%
\begin{figure}[!ht]
\centering
\includegraphics[height=6cm]{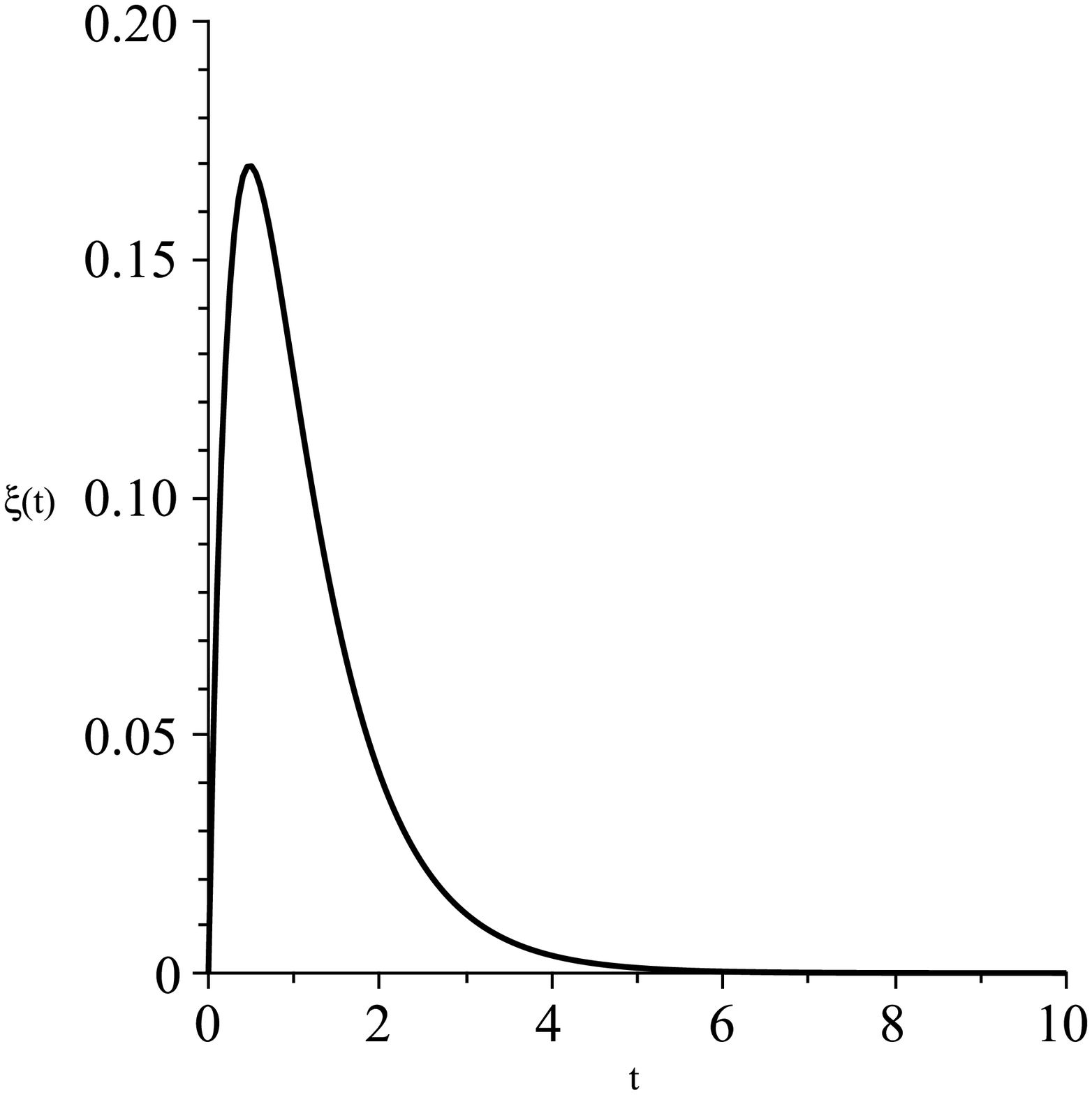}
\hfill
\includegraphics[height=6cm]{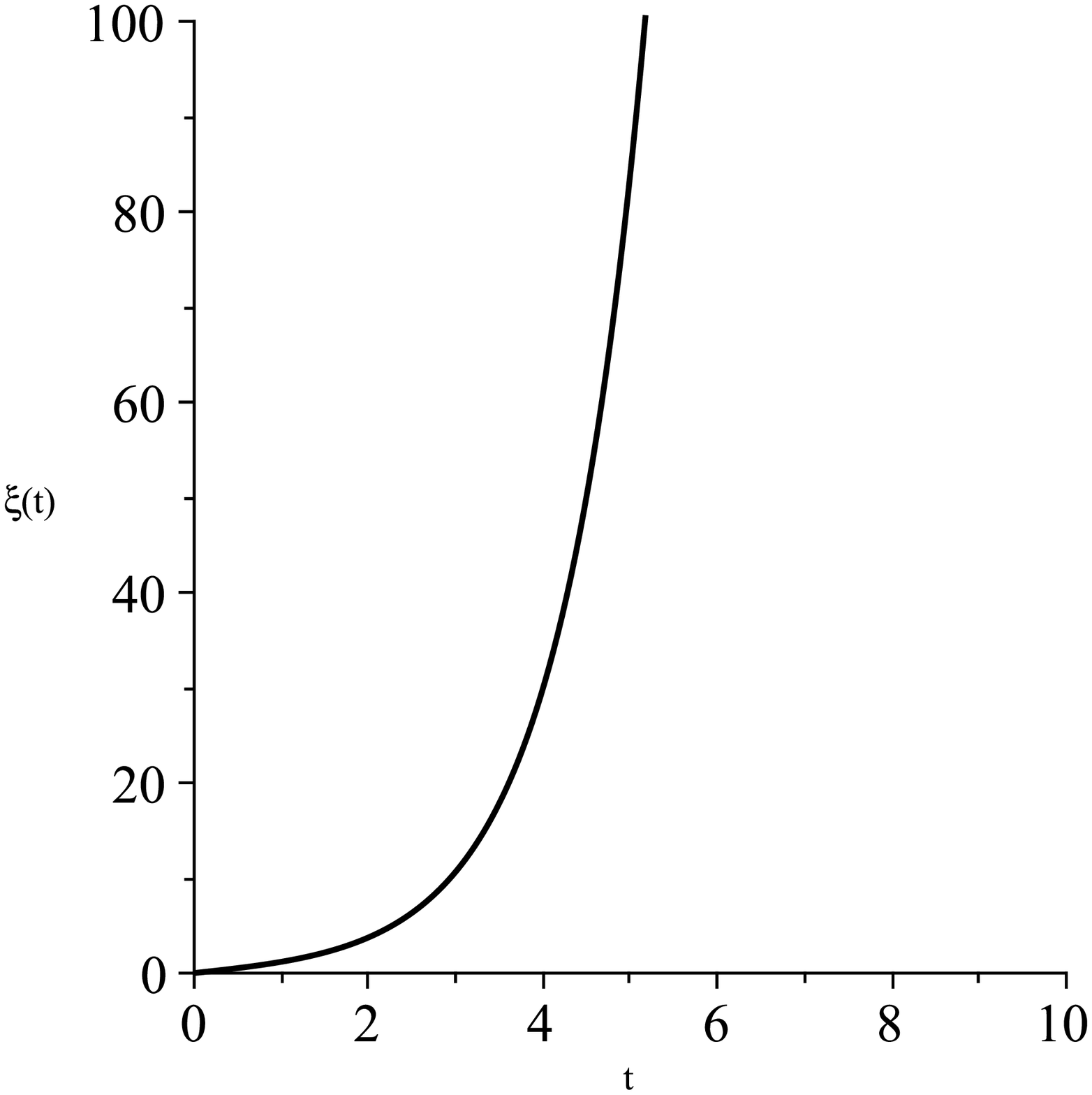}
\caption{Variation of the deviation vector field $\xi(t)$ given by \eqref{def} (left) and \eqref{JE} (right), for $a=4$, $b=0.5$.}
\end{figure}
%%%%%%%%%%%%%%%%%%%

The same kind of analysis can show how the deviation vector field variate in all four different regions explaining in this way the Table I given above.

\subsubsection{Limit cycles}

The following result is well known about the Brusselator (see for example \cite{Mu1993})

\begin{cor}
The fixed point $S(1,b/a)$ is

\begin{itemize}
 \item[$\circ$] an unstable node \index{unstable node}
if and only if $b>(\sqrt{a}+1)^2$, and
 \smallskip
 \item[$\circ$] an unstable spiral \index{unstable spiral}
if and only if $b<(\sqrt{a}+1)^2$.
\end{itemize}
Moreover, the point $b_c=a+1$ is a critical point in the sense
of Hopf, and for $b>b_c$ the system exhibits a
limit cycle behavior. \index{limit cycle}
\end{cor}

Indeed, it can be shown that in this case there is a confined closed set that contains the fixed point $S(1,\frac{b}{a})$ in its interior. From the Poincar\'e--Bendixon theorem, it follows that the system exhibits a limit cycle.

%%%%%%%%%%%%%%%%%%%%%%%%%%%%%%%%%%%%%%%%%%%%
%\begin{figure}[!ht]
%\centering
%\includegraphics[height=7cm, angle=270]{ymsriSa15.eps} %ymsriSab3.eps
%\hfill
%\includegraphics[height=7cm, angle=270]{ymsriSa16.eps} %ymsriSabnum.eps
%\caption{The Brusselator. Limit cycle behavior around an unstable spiral fixed point of the solutions of the Brusselator (left), numerical solution integrating the equations for $a=1$, $b=3$ (right).}
%\end{figure}
%%%%%%%%%%%%%%%%%%%%%%%%%%%%%%%%%%%%%%%%%%%%

\bigskip

We have seen that  if the fixed point $S(1,b/a)$ belongs to one of the regions $A$ or $B$, then it must behave as a limit cycle.

From Table I one can easily see that the Jacobi stability of the limit cycle changes when moving from the region A to B.

The usual bifurcation analysis does not give any information about the
stability of the periodic solutions. We are going to estimate the
stability of such a solution by means
of Jacobi stability.

First, we remark that because of $tr A>0$, and $det A >0$,
from Corollary 3.5 it follows:

\noindent{\bf Proposition 4.2}

\begin{itemize}
\item {\it If the fixed point $S(1,\frac{b}{a})$ is an
unstable spiral\index{stable spiral},
then it is in the Jacobi stability region.
Conversely, if it is in the Jacobi stability region, and $b>a+1$, then
it is an unstable spiral.}
\item {\it If the fixed point $S(1,\frac{b}{a})$ is an unstable node,
then it is in the Jacobi instability region.  Conversely, if it is in
the Jacobi instability region, and $b>a+1$, then it is  an
unstable node.}
\end{itemize}
This can also be seen directly from
$4P_1^1|_{(1,\frac{b}{a})}=\Delta = (tr A)^2 - 4 \, det A$.

 Therefore, in an open region around $S$, the transient trajectories are
Jacobi stable, so we can conclude that in this region the system is
robust\index{robust} and it becomes fragile outside it. This
conclusion is in accord with the usual
stability analysis done with Floquet theory.

The next sections introduce various physically motivated dynamical systems to which the previously discussed techniques can be applied. While following the linear stability results are mainly know, most of the remaining analysis is new.

\section{Newtonian polytropes -- Lane-Emden equation}\label{4}

It is well known that the Lane-Emden or Emden-Fowler equation can be rewritten as an autonomous system of equations by introducing a new set of variables. The Lane-Emden equations appear very naturally in the study os compact astrophysical objects in Newtonian gravity.

\subsection{The Lane-Emden equation and linear stability}

The properties of the stable Newtonian compact astrophysical objects can be completely described by the gravitational structure equations, which are given by:
\begin{equation}
  \frac{dm(r)}{dr}=4\pi \rho r^{2},
  \label{5}
\end{equation}
\begin{equation}
  \frac{dp}{dr}=-\frac{Gm(r)}{r^{2}}\rho,
  \label{6}
\end{equation}
where $\rho \geq 0$ is the density, $p \geq 0$ is the thermodynamical pressure, $m(r)$ denotes the mass inside radius $r$, satisfying the condition $m(r)\geq 0,\ \forall r \geq 0$. By eliminating the mass between Eqs.~(\ref{5}) and (\ref{6}) we obtain the equation \cite{Ho04}
\begin{equation}
  \frac{1}{r^{2}}\frac{d}{dr}\left(\frac{r^{2}}{\rho }\frac{dp}{dr}\right)
  =-4\pi G\rho.
  \label{7}
\end{equation}
To close this structure equation, one assumed an equation of state to specify the matter content, $p=p(\rho)$. We assume that the equation of state is continuous and it is sufficiently smooth for all $p>0$. A physically meaningful solution of Eqs.~(\ref{5})-(\ref{7}) is possible only when boundary conditions are imposed. We require that the interior of any matter distribution be free of singularities, which imposes the condition $m(r)\rightarrow 0$ as $r\rightarrow 0$. At the center of the star the other boundary conditions for Eqs.~(\ref{5})--(\ref{7}) are $p(0)=p_{c}$, $\rho(0)=\rho_{c}$, where $\rho_{c}$ and $p_{c}$ are the central density and pressure, respectively. The radius $R$ of the star is determined by the vanishing pressure surface $p(R)=0$.

An isotropic bounded fluid distribution for which the pressure and the density are related by a power law of the form
\begin{equation}
  p=K\rho ^{1+1/n},
\end{equation}
where $K$ and $n$ are constants, plays an important role in astrophysics. $n>0$ is called the polytropic index. It is convenient to represent the density in terms of a new dimensionless variable $\theta$ defined by
\begin{equation}
  \rho =\rho _{c}\theta ^{n},
\end{equation}
giving for the pressure $p=K\rho _{c}^{1+1/n}\theta ^{n+1}$. By rescaling the radial coordinate as $r=\alpha \xi $, $\alpha =\left[ \left( n+1\right) K\rho _{c}^{1/n-1}/4\pi G\right] ^{1/2}$, $n\neq -1,\pm \infty $, for polytropic stars Eq.~(\ref{7}) takes the form of the Lane-Emden equation of index $n$,
\begin{equation}
  \frac{1}{\xi ^{2}}\frac{d}{d\xi }
  \left(\xi ^{2}\frac{d\theta }{d\xi }\right) = -\theta ^{n}.
  \label{LE}
\end{equation}

The initial conditions for the Lane-Emden equation are $\theta(0) =1$ and $\theta'(0)=0$, respectively, where the prime denotes the derivative with respect to the variable $\xi$. By introducing a set of new variables $(w,t)$ defined as \cite{Ho87}
\begin{equation}
  \theta(\xi) = B\xi^{2/(1-n) }w(\xi),\qquad
  \xi = e^{-t},
\end{equation}
where $B>0$ is a constant, the Lane-Emden equation is transformed into the following second order differential equation,
\begin{equation}
  \frac{d^{2}w}{dt^{2}}+\frac{5-n}{n-1}\frac{dw}{dt}+
  \frac{2\left( 3-n\right)}{\left( n-1\right) ^{2}}w+
  B^{n-1}w^{n}=0,\qquad n\neq 1,\pm \infty.
  \label{u}
\end{equation}

In the following we will restrict the range of the polytropic index $n$ to the range $n\in \left( 1,+\infty \right) $. From a mathematical point of view Eq.~(\ref{u}) is a second order non-linear differential equation of the form $d^{2}w/dt^{2}+2G^{1}\left(w,dw/dt\right) =0$, with
\begin{equation}
  G^{1}\left( w,\frac{dw}{dt}\right) =\frac{1}{2}\left[ \frac{5-n}{n-1}
  \frac{dw}{dt}+\frac{2\left( 3-n\right) }{\left( n-1\right)^{2}}w+
  B^{n-1}w^{n}\right].
\end{equation}
The function $G^{1}$ has the properties $G^{1}(0,0)=0$ and $G^{1}(w,0)\neq 0$ for $w\neq 0$, respectively.

In order to study the stability of the equilibrium points of Eq.~(\ref{u}), we rewrite it in the form of an autonomous dynamical system, by introducing a new variable $q=dw/dt$, so that the Lane-Emden equation is equivalent to the following first order autonomous system of equations
\begin{equation}
  \frac{dw}{dt}=q,\qquad
  \frac{dq}{dt}=-2G^{1}(w,q).
  \label{syst}
\end{equation}
The critical points of this dynamical system are given by $q=0$, and by the
solutions of the equation
\begin{equation}
  G^{1}(w,0) = 0.
  \label{eqal}
\end{equation}

Therefore for the critical points $X_{n}=(w_{0},q_{0})$ of the system (\ref{syst}) we find the values
\begin{equation}
  X_{0}=\left( 0,0\right),\qquad n \in \left( 1,+\infty \right),
  \label{critp1}
\end{equation}
\begin{equation}
  %\label{critp2}
  X_{n}=\left( \frac{1}{B}\left[ \frac{2\left( n-3\right)}
  {\left( n-1\right)^{2}}\right]^{1/\left( n-1\right) },0\right),
  \qquad n \in (3,\infty),
  \label{critp3}
\end{equation}
\begin{equation}
  X_{in}=\left( \frac{i^{2/(n-1)}}{B}\left[ \frac{2\left(n-3\right)}
  {\left(n-1\right)^{2}}\right]^{1/\left( n-1\right) },0\right),
  \qquad n \in (1,3).
  \label{critp4}
\end{equation}

\subsection{Jacobi stability analysis of Newtonian polytropic fluid spheres}

Following our previous work~\cite{BoHa09} we denote $x:=w$ and $\ y:=dx/dt=dw/dt$. Then Eqs.~(\ref{u}) can be written as
\begin{equation}
  \frac{d^{2}x}{dt^{2}}+2G^{1}\left( x,y\right) =0,
  \label{jac2}
\end{equation}
where
\begin{equation}
  G^{1}\left( x,y\right) =\frac{1}{2}\left[ \frac{5-n}{n-1}y+
  \frac{2\left( 3-n\right) }{\left( n-1\right)^{2}}x+B^{n-1}
  \left( x\right)^{n}\right],
  \label{G1}
\end{equation}
respectively, which can now be studied by means of the KCC theory. The following treatment can also be found in one of our previous works~\cite{BoHa09}. As a first step in the KCC stability analysis of the Newtonian polytropes we obtain the nonlinear connections $N_{1}^{1}=\partial G^{1}/\partial y$, associated to Eqs.~(\ref{jac2}), and which is given by $N_{1}^{1}(x,y) = (5-n)/2(n-1)$. For the Lane-Emden equation the associated Berwald connection $G_{11}^{1} = \partial N_{1}^{1}/\partial y$ is identically equal to zero $G_{11}^{1} \equiv 0$. Finally, the second KCC invariant, or the deviation curvature tensor $P_{1}^{1}$, defined as
\begin{equation}
  P_{1}^{1}=-2\frac{\partial G^{1}}{\partial x}-2G^{1}G_{11}^{1}+
  y\frac{\partial N_{1}^{1}}{\partial x}+N_{1}^{1}N_{1}^{1}+
  \frac{\partial N_{1}^{1}}{\partial t},
\end{equation}
is given by
\begin{equation}
  P_{1}^{1}(x,y) =\frac{1}{4}-nB^{n-1}(x)^{n-1},
\end{equation}
or, in the initial variables, by
\begin{equation}
  P_{1}^{1}(w,q) =\frac{1}{4}-nB^{n-1}w^{n-1}.
\end{equation}

Evaluating $P_{1}^{1}(w,q)$ at the critical points $X_{n}$, given by Eqs.~(\ref{critp1})--(\ref{critp3}), we obtain
\begin{equation}
  P_{1}^{1}\left(X_{0}\right) = \frac{1}{4}>0,
  \qquad n \in (1,\infty),
\end{equation}
\begin{equation}
  P_{1}^{1}\left(X_{in}\right) = \frac{9n^{2}-26n+1}{4(n-1)^{2}},
  \qquad n \in (1,3).
\end{equation}
\begin{equation}
  P_{1}^{1}\left(X_{n}\right) = \frac{-7n^{2}+22n+1}{4(n-1)^{2}},
  \qquad n\in (3,\infty).
\end{equation}

\begin{figure}[!ht]
\centering
\includegraphics[width=0.48\textwidth]{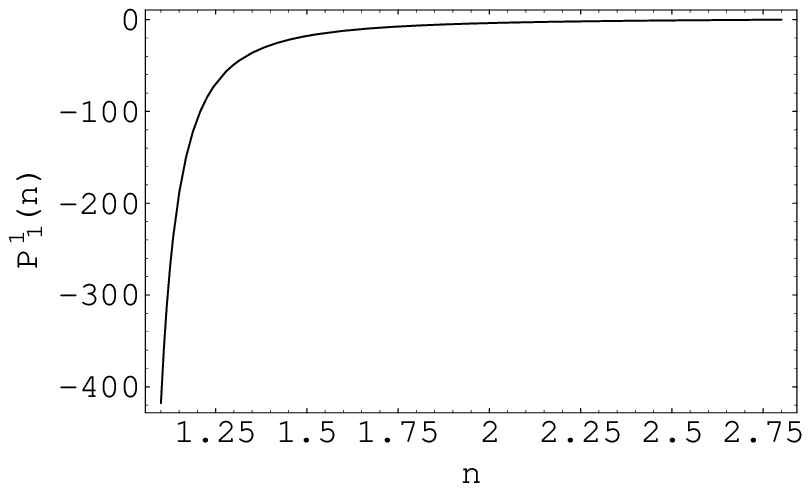}\hfill
\includegraphics[width=0.48\textwidth]{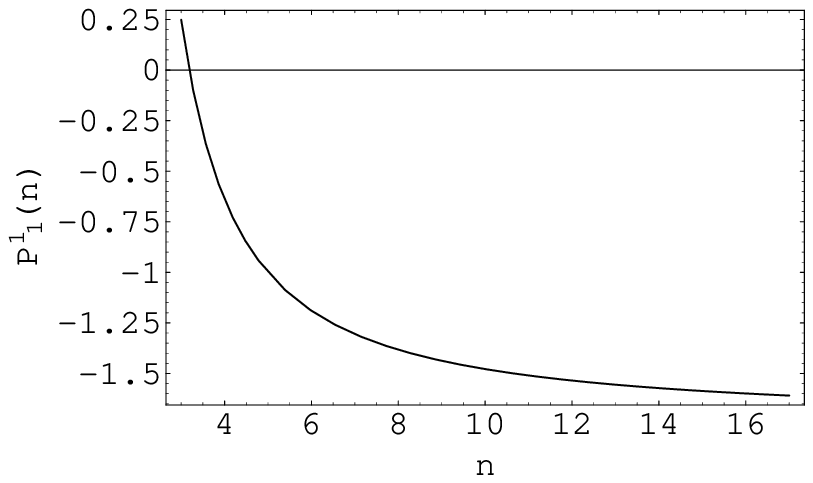}
\caption{Left: The deviation curvature tensor $P_{1}^{1}\left( X_{in}\right)$ as a function of $n\in(1,3)$. Right: The deviation curvature tensor $P_{1}^{1}\left( X_{n}\right)$ as a function of $n\geq 3$.}
\label{fig1}
\end{figure}

The behavior of the functions $P_{1}^{1}(n)$ is represented in Fig~\ref{fig1}.

In the initial variables the deviation curvature tensor $P_{1}^{1}$ is given by
\begin{equation}
  P_{1}^{1}(\xi,\theta) =\frac{1}{4}-n\xi^{2}\theta^{n-1}.
\end{equation}
$P_{1}^{1}$ can be expressed in a simple form with the use of Milne's homological variables $(u,v)$, defined as \cite{Ho87,Ho04}
\begin{equation}
  u=-\frac{\xi \theta ^{n}}{\theta ^{\prime }},\qquad
  v=-\frac{\xi \theta ^{\prime }}{\theta},
\end{equation}
with the use of whom we obtain
\begin{equation}
  P_{1}^{1}(u,v) = \frac{1}{4} - n\,uv.
\end{equation}

From a physical point of view $u(r)$, defined as
\begin{equation}
  u(r)=d\ln m(r)/d\ln r=3\rho (r)/\bar{\rho}(r),
\end{equation}
is equal to three times the density of the star at point $r$ divided by the mean density of matter $\bar{\rho}(r)$ contained in the sphere of radius $r$. The variable $v(r)$, defined as $v(r)=-d\ln p(r)/d\ln r=(3/2)\left[ Gm(r)/r\right] /\left[ 3p(r)/2\rho (r)\right] $, is $3/2$ of the ratio of the absolute value of the potential energy $\left|E_{g}\right| =Gm(r)/r$ and the internal energy per unit mass $E_{i}=(3/2)p/\rho $, so that $v(r)=(3/2)\left| E_{g}\right| /E_{i}$ \cite{Ho04}. In terms of these physical variables the deviation curvature tensor is given by
\begin{equation}
  P_{1}^{1}(r)=\frac{1}{4}-\frac{3n}{2}
  \frac{\rho (r)}{\bar{\rho}(r)}\frac{\left| E_{g}\right| }{E_{i}}.
\end{equation}
The condition of the Jacobi stability of the trajectories of the dynamical system describing a Newtonian polytropic star, $P_{1}^{1}<0$, can be formulated as
\begin{equation}
\frac{E_{i}}{\left| E_{g}\right| }<6n\frac{\rho (r)}{\bar{\rho}(r)}.
\end{equation}

\section{The general relativistic static fluid sphere}
\label{sec5}

The properties of an isotropic compact general relativistic object can be completely described by the gravitational structure equations, which are given by \cite{Co77}:
\begin{equation}
  \frac{dm}{dr}=4\pi \rho r^{2},
  \label{5_1}
\end{equation}
\begin{equation}
  \frac{dp}{dr}=-\frac{\left(\rho +p\right) \left( m+4\pi r^{3}p\right)}
  {r^{2}\left(1-\frac{2m}{r}\right)},
  \label{6_1}
\end{equation}
where $m(r)$ is the mass inside radius $r$ satisfying the condition $m(r)\geq 0,\forall r\geq0$. To close the field equations an equation of state $p=p(\rho)$ must also be given. We assume that the barotropic equation of state is continuous and it is sufficiently smooth for $p>0$. In the Newtonian limit Eq.~(\ref{6_1}) reduces to the expression given by Eq.~(\ref{6}).

A physically meaningful solution of Eqs.~(\ref{5_1})--(\ref{6_1}) is only possible when boundary conditions are imposed. We require that the interior of any matter distribution be free of singularities, which imposes the condition $m(r)\rightarrow 0$ as $r\rightarrow 0$. At the center of the star  the other boundary conditions for Eqs.~(\ref{5_1})--(\ref{6_1}) are
\begin{equation}
  \rho(0) = \rho_{c},\qquad
  p(0) = p_{c},
\end{equation}
where $\rho_{c}$ and $p_{c}$ are the central density and pressure, respectively. The radius $R$ of the star is determined by the boundary condition $p(R)=0$.  Depending on the form of the equation of state of the matter several fluid star models can be considered.

By introducing the set of the homologous variables $P$, $\mu $, $M$ and $\theta $, defined as
%%%
%%% I rewrote
%%% u = M
%%% v = mu
%%%
\begin{equation}\label{variab}
  \pi(v) = 4\pi r^2\, p(\rho(v)), \quad
  v = 4\pi r^2 \rho, \quad
  u = m/r, \quad
  t = \ln(r/R),
\end{equation}
the structure equations of the star take the form
\begin{eqnarray}
  \label{eqff1}
  \frac{du}{dt} &=& v - u,\\
  \label{eqff2}
  \frac{d\pi}{dt} &=& 2\pi - \frac{(v+\pi(v))(u+\pi(v))}{1-2u}.
\end{eqnarray}
On the boundary of the star, where $r=R$ we have $t=0$. At the center of the star, where $r\rightarrow 0$, we have $t\rightarrow\infty$. In order to close the system of equations (\ref{eqff1})--(\ref{eqff2}), the equation of state of the matter must be given in the form $\pi=\pi(v)$.

\subsection{Compact objects and linear stability}

As we have seen in Section~\ref{kcc}, the stability properties of a dynamical system of second order differential equations can be investigated by using the KCC geometric approach. The KCC theory gives the possibility of systematically investigating the Jacobi stability properties of the general relativistic compact objects in a rigorous framework.  As a first example of the application of the KCC theory to the fluid sphere model we consider the case of in which the equation of state of the fluid is given by a linear relation between the pressure and the energy density, so that
\begin{equation}
  p=(\gamma-1)\rho,\qquad 1<\gamma \leq 2.
\end{equation}

The equation of state of the homologous variables is $\pi(v)=(\gamma-1)v$, and the sound speed in the medium is given by $c_s^2=\gamma -1$. The structure equations of the isotropic star with a linear barotropic equation of state take the form of an autonomous system \cite{Co77}, given by
\begin{eqnarray}
  \label{is1}
  \frac{du}{dt} &=& v - u,\\
  \label{is2}
  \frac{dv}{dt} &=& 2v - \frac{v}{(\gamma-1)}
  \left(2 - \gamma v - \frac{5\gamma-4}{\gamma-1}u \right),
\end{eqnarray}

By eliminating $\mu $ between these equations we obtain the second order differential equation for $M$ in the form
\begin{equation}
  \label{eqMstab}
  \frac{d^2u}{dt^2}+\frac{du}{dt}-
  \frac{(u+du/dt)}{1-2u}\left[2-\frac{\gamma^2+4\gamma-4}{\gamma-1}u-\gamma\frac{du}{dt}\right]=0.
\end{equation}

%%%%%%%%%%%%%%%%%%%%%%%%%%%%%%%%%%%%%%%%%%%%
% Linear stability analysis
%%%%%%%%%%%%%%%%%%%%%%%%%%%%%%%%%%%%%%%%%%%%%%

We give here a short review of the linear analysis of the autonomous system given by  Eqs.~(\ref{is1}) and (\ref{is2}), respectively. One can find a similar analysis in \cite{Co77}. There are two steady states for this dynamical system, namely
\begin{eqnarray}
  \label{is16}
  (u_0,v_0) &=& (0,0),\\
  \label{is26}
  (u_1,v_1) &=& \Bigl(\frac{2(\gamma-1)}{\gamma^2+4\gamma-4},
  \frac{2(\gamma-1)}{\gamma^2+4\gamma-4}\Bigr),
\end{eqnarray}
respectively.

Let us remark two things about these steady states. Firstly, one can see that $(u_0,v_0)$ is not desirable because of the conditions $u(t)>0$, $v(t)>0$ imposed on the system by (\ref{variab}), and therefore the only acceptable steady state is $(u_1,v_1)$. One should also remark that for $u=1/2$, the system has a singularity, so actually we are concerned only with $0<u<1/2$. Secondly, for $\gamma=-2\pm 2\sqrt{2}$, the denominator of  $(u_1,v_1)$ vanishes. However, taking into account the condition $1<\gamma\leq 2$, we remark that $(u_1,v_1)$ is well defined in our case.

The Jacobian matrix of Eqs.~(\ref{is1}) and (\ref{is2}) is given by
\begin{equation}
  J=\begin{pmatrix}
  \frac{2\gamma(\gamma-1)\mu-(5\gamma-4)M-2(\gamma-1)}{(\gamma-1)(2M-1)} &
  -\frac{\gamma\mu [1+2(\gamma-1)\mu]}{(\gamma-1)(2M-1)^2}\\
  1 & -1
  \end{pmatrix},
\end{equation}
and by evaluating it at $(u_1,v_1)$ we obtain
\begin{equation}
  A_1=J(u_1,v_1)=
  \begin{pmatrix}
    -\frac{2(\gamma-1)}{\gamma} & -\frac{2(5\gamma-4)}{\gamma^2} \\
    1 & -1
  \end{pmatrix}.
\end{equation}

The eigenvalues of the matrix $A_1$ are the solutions of the characteristic equation
\begin{equation}
  \label{char1}
  \lambda^2-\lambda \cdot {\rm tr}(A_1)+\det A_1=0,
\end{equation}
where
\begin{equation}
  {\rm tr}(A_1)=-\frac{3\gamma-2}{\gamma},\qquad
  \det A_1=\frac{2(\gamma^2+4\gamma-4)}{\gamma^2},
\end{equation}
are the trace and determinant of the matrix $A_1$. Moreover, the discriminant of the characteristic equation (\ref{char1}) is given by
\begin{equation}
  \Delta=({\rm tr}(A_1))^2-4\det
  A_1=\frac{\gamma^2-44\gamma+36}{\gamma^2}.
\end{equation}

\begin{figure}[!ht]
\centering
\includegraphics[width=0.48\textwidth,angle=270]{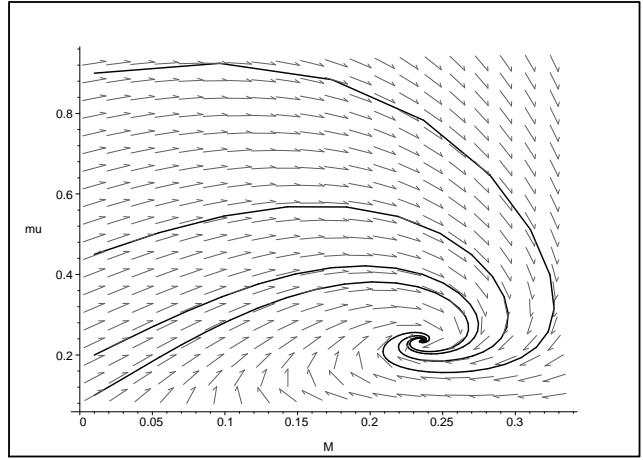}
\caption{The trajectories of the autonomous system (\ref{is1}), (\ref{is2}) for $\gamma=1.5$.}
\label{fig1_1}
\end{figure}

Remark that for $1<\gamma\leq 2$ we always have $\Delta<0$, i.e.~the eigenvalues of the matrix $A_1$ are complex. Moreover, since ${\rm tr}(A_1)<0$ for the given range of $\gamma$, it follows that the steady state $(u_1,v_1)$ is a {\it stable spiral}, as shown in Fig.~\ref{fig1_1}.

Remark that for  $1<\gamma\leq 2$ there are no bifurcations points, i.e.~the stable spirals are very robust.

%%%%%%%%%%%%%%%%%%%%%%%%%%%%%%%%%%%%%%%%%%%%
% Jacobi stability analysis
%%%%%%%%%%%%%%%%%%%%%%%%%%%%%%%%%%%%%%%%%%%%%%

\subsection{Jacobi stability analysis}

The study of the Jacobi stability analysis of the general relativistic fluid sphere with a linear barotropic equation of state is based on the direct study of the second order differential equation given by Eq.~(\ref{eqMstab}). In order to simplify the notation we introduce the new variables $x:=u$ and $y:=dx/dt =du/dt$. Therefore Eq.~(\ref{eqMstab}) can be written as
\begin{equation}
  \frac{d^2x}{d\theta ^2}+y-\frac{(x+y)}{1-2x}
  \left[2-\frac{\gamma^2+4\gamma-4}{\gamma-1}x-\gamma y\right]=0.
\end{equation}
In other words, we can express the system of the structure equations of the compact general relativistic star with a linear barotropic equation of state, given by Eqs.~(\ref{is1}) and (\ref{is2}), respectively, by the SODE
\begin{equation}
  \label{jac1}
  \frac{d^2x}{dt^2} + 2G^1(x,y) = 0,
\end{equation}
where
\begin{equation}
  2G^1(x,y) = y-\frac{(x+y)}{1-2x}\left[2-\frac{\gamma^2+4\gamma-4}{\gamma-1}x-\gamma y\right].
\end{equation}

As a first step in the KCC stability analysis of the general relativistic fluid sphere with linear barotropic equation of state we obtain the nonlinear connection $N_{1}^{1}$ associated to Eq.~(\ref{jac1}), and which is given by
\begin{equation}
  N_{1}^{1} = \frac{\partial G^{1}}{\partial y} =
  \frac{\left(\gamma -1\right)\left(2\gamma y-1\right)+
  \left(2\gamma ^2+\gamma -2\right)x}{2\left(\gamma -1\right)
  \left(1-2x\right)}.
\end{equation}
The Berwald connection can be obtained as
\begin{equation}
  G_{11}^{1} = \frac{\partial N_{1}^{1}}{\partial y} =
  \frac{\gamma }{1-2x}.
\end{equation}
Finally, the second KCC invariant or the deviation curvature tensor $P_{1}^{1}$, defined as
\begin{equation}
  P_{1}^{1} = -2\frac{\partial G^{1}}{\partial x}-
  2G^{1}G_{11}^{1}+y\frac{\partial N_{1}^{1}}{\partial x}+
  N_{1}^{1}N_{1}^{1},
\end{equation}
reads now
\begin{equation}
  P_1^{1} = \frac{(7\gamma-6)^2x^2-
  2(\gamma-1)(2\gamma^2+21\gamma -18)x-
  2\gamma(\gamma-1)(2\gamma-1)y+9(\gamma-1)^2}
  {4(\gamma-1)^2(1-2x)^2}.
\end{equation}
Equivalently, this can be written in the variables $(u,v)$ as follows
\begin{equation}
  P_1^{1} = \frac{(7\gamma-6)^2 u^2-4(\gamma-1)(11\gamma-9)u-
  2\gamma(\gamma-1)(2\gamma-1)v+9(\gamma-1)^2}
  {4(\gamma-1)^2(1-2u)^2}.
\end{equation}
Evaluating now $P_1^{1}(u,v)$ at the steady state $(u_1,v_1)$, we obtain a rather simple form of this function. Equivalently, in the variables $(u_1,v_1)$ this can be written as follows
\begin{equation}
  P_1^{1}(u_1,v_1)=\frac{\gamma^2-44\gamma+36}{4\gamma^2}.
\end{equation}

\begin{figure}[!ht]
\centering
\includegraphics[width=0.48\textwidth]{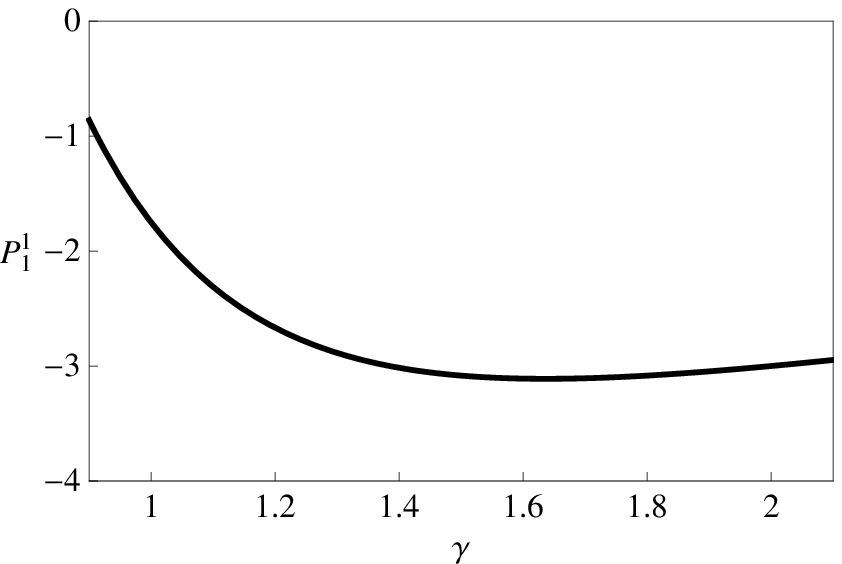}\\
\includegraphics[width=0.48\textwidth]{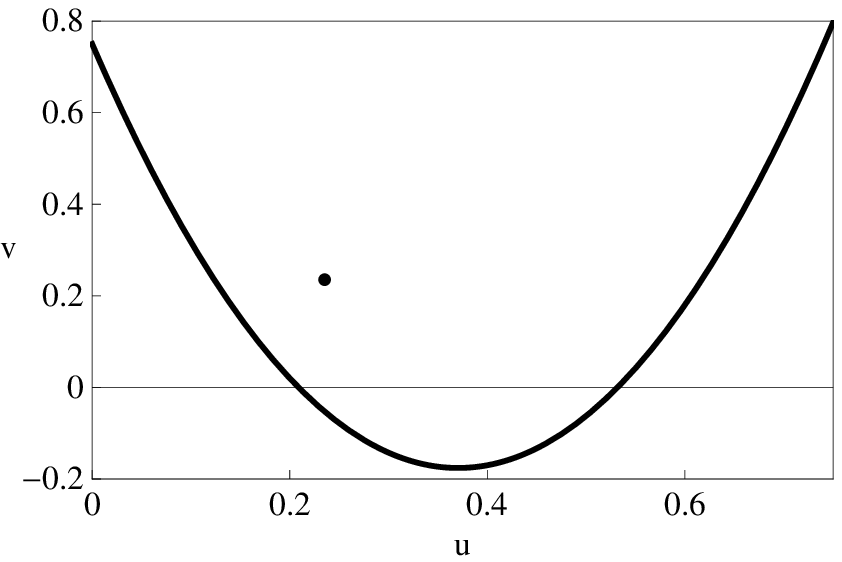}
\caption{Left: The graph of the function $P_1^{1}(u_1,v_1)$. Right: The graph of the function $P_1^{1}(u,v)$ for $\gamma=1.5$. The steady state $(u_1,v_1)$ denoted by a point in the graph, is always in the Jacobi stability region.}
\label{fig3_1}
\end{figure}

We remark that, for $1<\gamma\leq 2$, we have $P_1^{1}(u_1,v_1)<0$ (see Fig.~\ref{fig3_1}), and therefore, from the general KCC theory, it follows that the steady point $(u_1,v_1)$ is Jacobi stable without any bifurcation points (see Fig.~\ref{fig3_1}).

\begin{figure}[!ht]
\centering
\includegraphics[width=0.48\textwidth,angle=270]{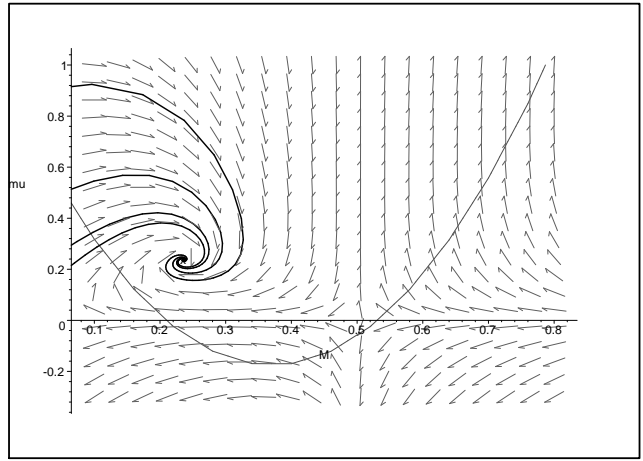}
\caption{The trajectories of the system are always attracted in the interior of the Jacobi stability region (we show the graph for $\gamma=1.5$).}
\label{fig4_1}
\end{figure}

Moreover, we remark that the trajectories of the system (\ref{is1}), (\ref{is2}) are always attracted to the steady state $(u_1,v_1)$ in the interior of the Jacobi stability region. In other word, even they start outside the Jacobi stability region, inevitable they will end by being Jacobi stable in the final (see
Fig.~\ref{fig4_1}).

The condition for the Jacobi stability of the star with a linear barotropic equation of state, $P_1^{1}(u,v)<0$, can be written in the initial physical variables $m(r),\rho(r)$ in the form of a restriction imposed on the physical parameters of the stars, and which is given by
\begin{equation}
  \left[\frac{m(r)}{r}\right]^2-\frac{4\left(\gamma -1\right)\left(11\gamma -1\right)}{\left(7\gamma -6\right)^2}\frac{m(r)}{r}-\frac{8\pi \gamma \left(\gamma -1\right)\left(2\gamma -1\right)}{\left(7\gamma -6\right)^2}\rho r^2+\frac{9\left(\gamma -1\right)^2}{\left(7\gamma -6\right)^2}<0.
\end{equation}
This condition must be satisfied at all points $0 <r \leq R$ inside the star. By means of some simple algebraical transformations we can rewrite this condition as
\begin{equation}
  \label{condlin}
  \frac{m(r)}{r}<\frac{2\left(\gamma -1\right)\left(11\gamma -9\right)}{\left(7\gamma -6\right)^2}+\frac{\gamma -1}{7\gamma -6}\sqrt{\frac{8\pi \gamma \left(2\gamma -1\right)}{\gamma -1}\rho r^2+\frac{4\left(11\gamma -9\right)^2}{\left(7\gamma -6\right)^2}-9}.
\end{equation}
Due to assuming a linear equation of state for the matter, at the surface of the star the matter density and the pressure both vanishes, so that also $\rho(R)=0$. The total mass of the star is $M=m(R)$. By evaluating Eq.~(\ref{condlin}) at the vacuum boundary of the star we obtain the following restriction on the mass-radius ratio for barotropic stars with a linear equation of state
\begin{equation}
\frac{M}{R}<\frac{2\left(\gamma -1\right)\left(11\gamma -9\right)}{\left(7\gamma -6\right)^2}+\frac{\gamma -1}{7\gamma -6}\sqrt{\frac{4\left(11\gamma -9\right)^2}{\left(7\gamma -6\right)^2}-9}.
\end{equation}
When we, for example, choose $\gamma = 2$, then this equations results in $2M/R<9/8$ which can be combined with our previous assumption $M/R>1/2$ (note that this condition automatically follow from the quadratic inequality) to give the joint result
\begin{equation}
  \frac{1}{2} < \frac{2M}{R} < \frac{9}{8},
\end{equation}
which is a very nice physical result coming out of KCC theory. Note, however, that this condition is somewhat weaker than the well-known Buchdahl inequality $2M/R < 8/9$ which implies the absence of horizons in the matter part of the spacetime. This cannot be concluded from the KCC approach.

%%%
%%% BRANE WORLD MODELS
%%%

\section{Stability of the vacuum in brane world models}
\label{sec6}

Brane world models have become a popular modification of Einstein's theory of general relativity. In this model one considers a five dimensional spacetime, called the bulk spacetime, in which a single four-dimensional brane is embedded. The working assumption of these models is that the  matter is confined to the brane, and only gravity can probe the extra dimension. The action of such a system has been formulated by~\cite{SMS00,SMS001} and a variety of interesting results have been derived in the field of brane world gravity.

When one assumes the four-dimensional manifold to be static and spherically symmetric (analogously to the previous Section), then it is possible to derive a set of structure equations of the vacuum very similar to Eqs.~(\ref{5_1})--(\ref{6_1}). The properties of the vacuum on the brane are described by the following system of differential equations ~\cite{Ha03,Ma04,Ma041,HaSa08}
\begin{equation}
  \frac{dM_{U}}{dr}= 3\alpha r^{2}U,
  \label{e2}
\end{equation}
and
\begin{equation}
  \frac{dU}{dr}=-\frac{(2U+P)\left[2M+M_{U}-\frac{2}{3}\Lambda r^{3}+
  \alpha (U+2P) r^{3}\right] }{r^{2}
  \left( 1-\frac{2M}{r}-\frac{M_{U}}{r}-
  \frac{\Lambda }{3}r^{2}\right) }-2\frac{dP}{dr}-
  \frac{6P}{r},
\label{e1}
\end{equation}
respectively. In these equations, the quantity $M_U$ is referred to as the dark mass, describing the vacuum gravitational field, exterior to a massive body, in the brane world model. $U$ and $P$ are the dark radiation and dark pressure, respectively. $\Lambda$ is the cosmological constant. The `matter' source are the projections of the higher dimensional gravitational field onto the brane. The brane  is {\it apriori} in a vacuum state.

To close the system of equations, a supplementary functional relation between one of the unknowns $U$, $P$ and $M_{U}$ is needed. Generally, this equation of state is given in the form $P=P(U)$. Once this relation is known, Eqs.~(\ref{e2})--(\ref{e1}) give a full description of the geometrical properties of the vacuum on the brane.

In the following we will restrict our analysis to the case $\Lambda =0$. Then the system of equations~(\ref{e2}) and~(\ref{e1}) can be transformed to an autonomous system of differential equations by means of the transformations
\begin{equation}
  u = \frac{2M}{r}+\frac{M_{U}}{r},\qquad
  v = 3\alpha r^{2}U,\qquad
  \pi(v) = 3\alpha r^{2}P(U(v)),\qquad
  t =\ln r,
\label{transf1}
\end{equation}
where $v$ and $\pi$ are the `reduced' dark radiation and pressure, respectively.

With the use of the new variables given by~(\ref{transf1}),  Eqs.~(\ref{e2}) and~(\ref{e1}) become
\begin{eqnarray}
  \frac{du}{dt} &=& v - u,
  \label{aut1}\\
  \frac{dv}{dt} &=& -\frac{(2v+\pi)\left[u+(v+2\pi)/3 \right] }{1-q}-
  2\frac{d\pi}{dt} + 2v - 2\pi.
\label{aut2}
\end{eqnarray}

Eqs.~(\ref{e2}) and~(\ref{e1}), or, equivalently,~(\ref{aut1}) and~(\ref {aut2}), are called the structure equations of the vacuum on the brane \cite {Ha03}. In order to close the system of equations (\ref{aut1}) and~(\ref {aut2}) an `equation of state' $p=p(\mu) $, relating the reduced dark radiation and the dark pressure terms, is needed.

The structure equations of the vacuum on the brane can be solved exactly in two cases, corresponding to some simple equations of state of the dark pressure. In the first case we impose the equation of state $2\mu +p=0$. From Eq. (\ref{aut2}) we immediately obtain $\mu =Qe^{-2\theta }$, while Eq. (\ref{aut1}) gives $q\left( \theta \right) =-Qe^{-2\theta }+U_{0}e^{-\theta }$, where $Q$ and $U_{0}=2GM$ are arbitrary constants of integration. Therefore we obtain the vacuum brane solution
\begin{equation}
  U=-\frac{P}{2}=\frac{Q}{3\alpha }\frac{1}{r^{4}}.
\end{equation}
This solution was first obtained in \cite{Da00}, and therefore corresponds to an equation of state of the dark pressure of the form $P=-2U$. The second case in which the vacuum structure equations can be integrated exactly corresponds to the equation of state $\mu +2p=0$. Then Eq. (\ref{aut2}) gives $q=2/3$, and the corresponding solution of the gravitational field equations on the brane was derived in \cite{Ha03}, which corresponds to an equation of state of the dark pressure of the form $P=-U/2$.

\subsection{Linear stability analysis}
\label{linstab}

Since generally the structure equations of the vacuum on the brane cannot be solved exactly, in this Section we shall analyze them by using methods from the qualitative analysis of dynamical systems~\cite{BoHa07}, by closely following the approach of~\cite{boyce}. We consider the case in which the dark pressure is proportional to the dark radiation, $P=\gamma U$, where $\gamma $ is an arbitrary constant, which can take both positive and negative values. As we have seen, several classes of exact solutions of the vacuum gravitational field equations on the brane can be described by an equation of state of this form. In the reduced variables $u$ and $v$ the linear equation of state is $\pi=\gamma v $, and the structure equations of the gravitational field on the brane have the form
\begin{eqnarray}
  \frac{du}{dt} &=& v - u,
  \label{aut3a}\\
  \frac{dv}{dt} &=& \frac{2(1-\gamma)}{(1+2\gamma)} v -
  \frac{\gamma+2}{1+2\gamma}\frac{v \left[ u+\frac{1+2\gamma }{3}v \right]}{1-u}.
  \label{aut4a}
\end{eqnarray}

Let us firstly analyze the special case where $\gamma =-1/2$. Then, by virtue of the second Eq.~(\ref{aut4a}), we obtain two possible exact solutions, either $u=v=0$ or $u=v=2/3$. The first of these solutions correspond to the vanishing of all physical quantities, and therefore we can discard it as unphysical. The second case corresponds to an exact solution, which has already been discussed. Let us henceforth assume that $\gamma \neq -1/2$. Let us write the dynamical system in the form
\begin{equation}
  \frac{d\xi}{dt}=A\xi +B,
  \label{autuse}
\end{equation}
where we have denoted
\begin{equation}
  \xi = \begin{pmatrix}
    u \\ v
  \end{pmatrix}
  ,\quad A = \begin{pmatrix}
    -1 & 1 \\
    0 & 2(1-\gamma )/(1+2\gamma )
  \end{pmatrix}
  ,\quad B=
  \begin{pmatrix}
    0 \\
    \frac{\gamma+2}{1+2\gamma}\frac{v\left[u+\frac{1+2\gamma}{3}v\right]}{1-u}
\end{pmatrix}.
\end{equation}

The system of equations~(\ref{autuse}) has two critical points, $X_{0}=(0,0)$, and
\begin{equation}
  X_{\gamma }=\left( \frac{3(1-\gamma )}{\gamma ^{2}+\gamma +7},
  \frac{3(1-\gamma )}{\gamma ^{2}+\gamma +7}\right) .  \label{critp}
\end{equation}
For $\gamma =1$, the two critical points of the system coincide. Depending on the values of $\gamma $, these points lie in different regions of the phase space plane~$(u,v)$. Since the term $||B||/||\xi ||\rightarrow 0$ as $||\xi ||\rightarrow 0$, the system of equations~(\ref{autuse}) can be linearized at the critical point $X_{0}$. The two eigenvalues of the matrix $A$ are given by $r_{1}=-1$ and $r_{2}=2(1-\gamma )/(1+2\gamma )$, and determine the characteristics of the critical point $X_{0}$. For $\gamma \in (-\infty ,-1/2)\cup \lbrack 1,\infty)$ both eigenvalues are negative and unequal. Therefore, for such values of $\gamma $ the point $X_{0}$ is an improper asymptotically stable node.

If $\gamma \in (-1/2,1)$, we find one positive and one negative eigenvalue, which corresponds to an unstable saddle point at the point $X_0$. Moreover, since the matrix $dA/d\xi(X_0)$ has real non-vanishing eigenvalues, the point $X_0$ is hyperbolic. This implies that the properties of the linearized system are also valid for the full non-linear system near the point $X_0$. It should be mentioned however, that this first critical point is the less interesting one from a physical point of view, since it corresponds to the `trivial' case where both physical variables vanish.

The structure equations can also be solved exactly for the value $\gamma =-2$. In that case, the non-linear term $B$ in Eq.~(\ref{autuse}) identically vanishes, and the system of equations becomes a simple linear system of differential equations. For $\gamma =-2$ the two eigenvalues of $A$ are given by $r_{1}=-1$ and $r_{2}=-2$, and the two corresponding eigenvectors are linearly independent. The general solution can be written as follows
\begin{equation}
  \xi _{\gamma =-2}=(u_{0}+v_{0})
  \begin{pmatrix} 1 \\ 0 \end{pmatrix} e^{-t}+u_{0}
  \begin{pmatrix}-1 \\ 1 \end{pmatrix} e^{-t},
\end{equation}
where $u_{0}=u(0)$ and $v_{0}=v(0)$. One can easily transform this solution back into the radial coordinate $r$ form by using $t =\ln (r)$, thus obtaining
\begin{equation}
  \mu _{\gamma =-2}=\frac{\mu_{0}}{r^{2}},\qquad
  q_{\gamma =-2}=\frac{q_{0}}{r}+\mu_{0}
  \left(\frac{1}{r}-\frac{1}{r^{2}}\right) .
\end{equation}

Let us now analyze the qualitative behavior of the second critical point $X_{\gamma }$. To do this, one has to Taylor expand the right-hand sides of Eqs.~(\ref{aut3a}) and~(\ref{aut4a}) around $X_{\gamma }$ and obtain the matrix $\tilde{A}$ which corresponds to the system, linearized around $X_{\gamma }$. This linearization is again allowed since the resulting non-linear term, $\tilde{n}$ say, also satisfies the condition $||\tilde{n}||/||\xi ||\rightarrow 0$ as $||\xi ||\rightarrow X_{\gamma }$. The resulting matrix reads
\begin{equation}
  \tilde{A}=
  \begin{pmatrix}
    -1 & 1 \\
    \frac{3(-\gamma ^{2}+5\gamma -4)}{(2+\gamma )^{2}(1+2\gamma )} & \frac{%
      \gamma -1}{2+\gamma }
  \end{pmatrix},
  \label{ew}
\end{equation}
and its two eigenvalues are given by
\begin{equation}
  r_{\pm }=\frac{-3-6\gamma \pm \sqrt{16\gamma ^{4}+8\gamma ^{3}+132\gamma
      ^{2}-28\gamma -47}}{4\gamma ^{2}+10\gamma +4}.
\end{equation}
For $-0.5 < \gamma < 0.674865$ the argument of the square root becomes negative and the eigenvalues complex. Moreover, the values $\gamma=-1/2$ and $\gamma=-2$ have to be excluded, since the eigenvalues in Eq.~(\ref{ew}) are not defined in these cases. However, both cases have been treated separately above.

If $\gamma \in (-\infty ,-1/2)\cup (1,\infty )$, then $X_{\gamma }$ corresponds to an unstable saddle point and for $\gamma \in (0.674865,1)$ it corresponds to an asymptotically stable improper node. More interesting is the parameter range $\gamma \in (-1/2,0.674865)$, where the eigenvalues become complex, however, their real parts are negative definite. Hence, for those values we find an asymptotically stable spiral point at $X_{\gamma }$. Since this point is also a hyperbolic point, the described properties are also valid for the non-linear system near that point.

\begin{table}
\begin{center}
\begin{tabular}{|c|ccccccccc|}
\hline
 $\gamma$ & {$-\infty $} &  & -0.5 &  & 0.67 &  & 1 &  & +$\infty$ \\
\hline
{$r_{\pm}$} &  & real & \big| & complex & \big| & real & \big| & real &  \\
\hline &  &  & \big| & Re {$r_{\pm}<0$} & \big| &  & \big| &  &
\\ \hline {$r_{+}$} &  & + & \big| &  & \big| & -- & \big| & -- &
\\ \hline {$r_{-}$} &  & -- & \big| &  & \big| & -- & \big| & + &
\\ \hline
{$X_{\gamma}$} &  & Saddle & \big| & Stable spiral & \big| & Stable node & %
\big| & Saddle & \\ \hline
\end{tabular}
\end{center}
\caption{Linear stability of the stable point
$X_{\gamma}$ of the static brane world vacuum field equations.}\label{table1}
\end{table}

\begin{figure}[!ht]
\centering
\includegraphics[height=8cm,width=7.5cm,angle=270]{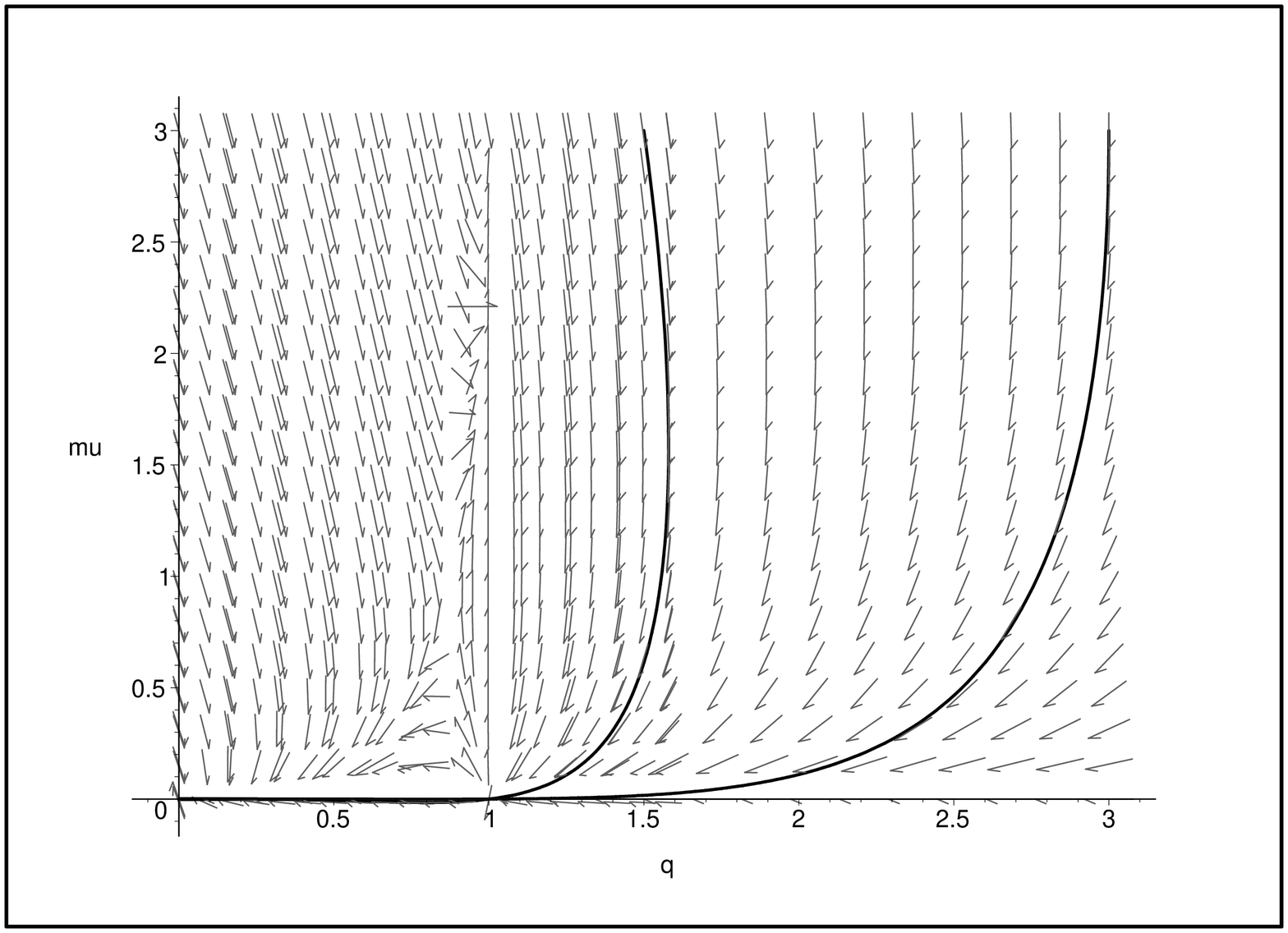}
\hfill
\includegraphics[height=8cm,width=7.5cm,angle=270]{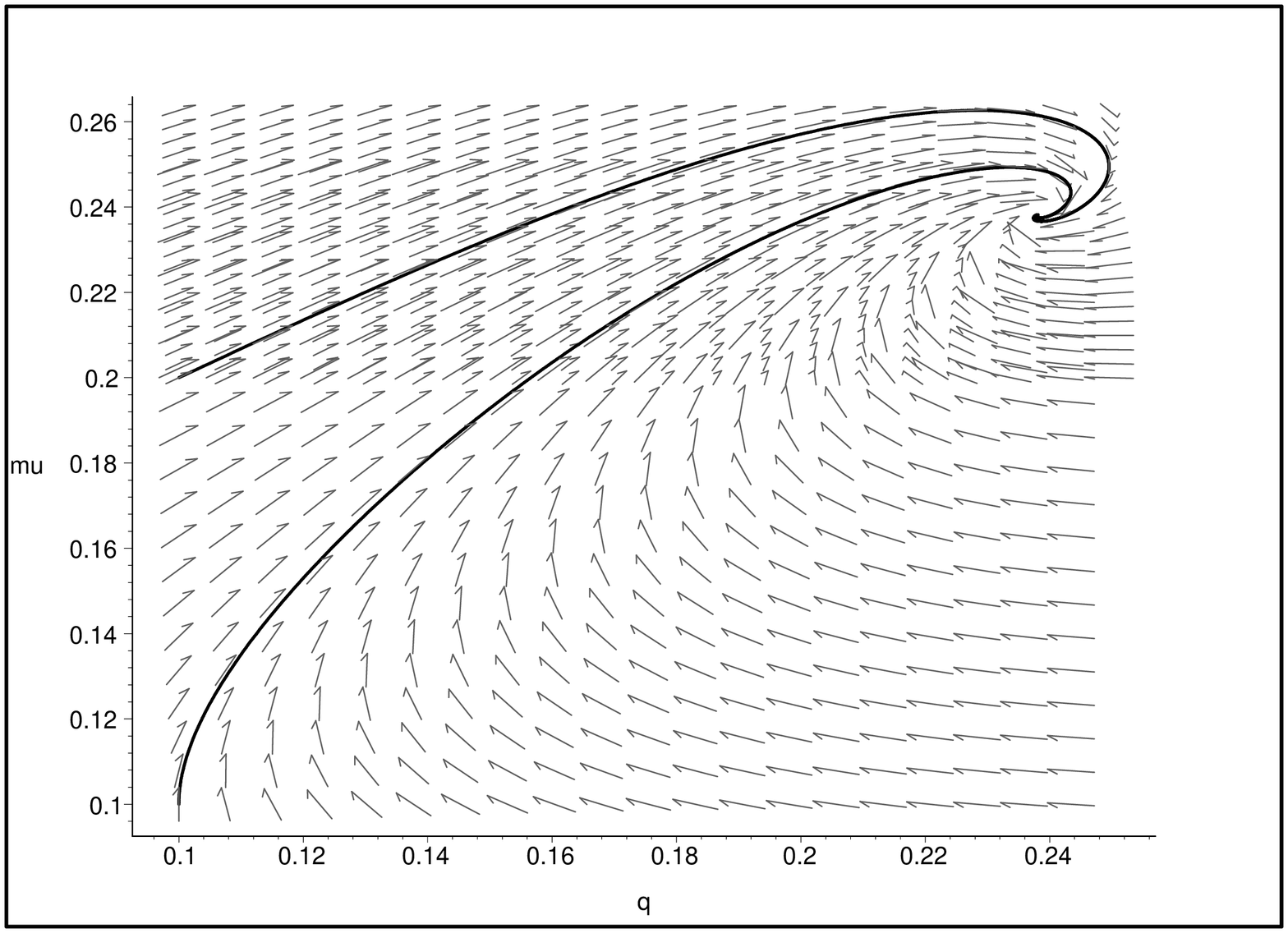}
\caption{Behavior of the trajectories of the structure equations
of the vacuum on the brane near the critical points for $\gamma
=-1$ (left figure) and $\gamma =0.4$ (right figure).}
\label{figf2}
\end{figure}

The behavior of the trajectories is shown, for $\gamma =-1$ and $\gamma =0.4$, in Fig.~\ref{figf2}. The figures show the attracting or repelling character of the steady states, respectively. The results of the linear stability analysis of the critical points $X_{\gamma}$ are summarized in Table~\ref{table1}.

\subsection{Jacobi stability analysis}

Since from Eq.~(\ref{aut3a}) we can express $v$ as $v=u+du/dt$, upon substitution in Eq.~(\ref{aut4a}) we obtain the following second order differential equation
% Am schimbat ecuatia pe 2 randuri ca sa scot overflow-ul. Cea veche
%este cu comment out. Daca nu va place, reveniti la varianta initiala
%
% I have also checked this equation by hand and it is correct
\begin{equation}
\begin{split}
  \frac{d^{2}u}{dt^{2}} & + \frac{1}{3(1+2\gamma)(1-u)}
  \Bigl[6(\gamma-1) u+ 2(\gamma^{2}+\gamma +7)u^{2}+
  3(4\gamma-1) du/dt\\
  & + (4\gamma^{2}+\gamma +13) u\,du/dt+
  (2\gamma^{2}+5\gamma+2)(du/dt)^{2} \Bigr]=0,
\end{split}
\label{jac11}
\end{equation}
which can now be studied by means of KCC theory.

By denoting $x=u$ and $du/dt=dx/dt=y$, Eq.~(\ref{jac11}) can be written as
\begin{equation}
\frac{d^{2}x}{d\theta ^{2}}+2G^{1}\left( x,y\right) =0,
\end{equation}
where
\begin{multline}
  G^{1}(x,y) = \frac{1}{6\left( 1+2\gamma \right) \left(1-x\right)}
  \Bigl[6\left( \gamma -1\right) x+2\left(
    \gamma ^{2}+\gamma +7\right) x^{2}+3\left( 4\gamma -1\right)y\\
    +\left( 4\gamma^{2}+\gamma +13\right) xy+\left( 2\gamma ^{2}+5\gamma +2\right) y^{2}\Bigr]
\end{multline}

As a first step in the KCC stability analysis of the vacuum field
equations on the brane we obtain the nonlinear connection
$N_{1}^{1}$ associated to Eq. (\ref{jac1}), and which is given by
\begin{equation}
  N_{1}^{1}=\frac{\partial G^{1}}{\partial y}=\frac{3\left( 4\gamma
    -1\right) +\left( 4\gamma ^{2}+\gamma +13\right) x+2\left( 2\gamma
    ^{2}+5\gamma +2\right) y}{6\left( 1+2\gamma \right) \left(
    1-x\right) }.
\end{equation}

%%%%%%%%%%%
%Am simplificat un 2 in formula care urmeaza
%%%%%%%%%%%%%
The Berwald connection can be obtained as
\begin{equation}
G_{11}^{1}=\frac{\partial N_{1}^{1}}{\partial y}=\frac{ 2\gamma
^{2}+5\gamma +2 }{3\left( 1+2\gamma \right) \left( 1-x\right) }.
\end{equation}

%%%%%%%%%%
%Am schimbat semnele in formula care urmeaza
%%%%%%%%%
Finally, the second KCC invariant or the deviation curvature tensor $%
P_{1}^{1}$, defined as
\begin{equation}
P_{1}^{1}=-2\frac{\partial G^{1}}{\partial x}-2G^{1}G_{11}^{1}+y\frac{%
\partial N_{1}^{1}}{\partial x}+N_{1}^{1}N_{1}^{1},
\end{equation}
reads now %%%%%%%%%%
%Am schimbat semnele in formula care urmeaza
%%%%%%%%%
\begin{multline}
  P_{1}^{1}(x,y) =\frac{27-2\left[ 61+57\gamma +4\gamma^{2}\left( 9+2\gamma \right) \right] x+3\left( 5+\gamma \right)^{2}x^{2}}{12\left( 1-x\right) ^{2}\left( 1+2\gamma \right) ^{2}}\\
    -\frac{2\left(2+\gamma \right) \left( 1+2\gamma \right) \left( 5+4\gamma \right) y}{12\left( 1-x\right) ^{2}\left( 1+2\gamma \right)^{2}}.
\end{multline}

Taking into account that $x=q$ and $y=\mu -q$, we obtain
$P_{1}^{1}$ in the
initial variables as %%%%%%%%%%
%Am schimbat semnele in formula care urmeaza
%%%%%%%%%
\begin{equation}
P_{1}^{1}\left( q,\mu \right) =\frac{27-6\left( 17+8\gamma
+2\gamma ^{2}\right) q+3\left( 5+\gamma \right) ^{2}q^{2}-2\left(
2+\gamma \right) \left( 1+2\gamma \right) \left( 5+4\gamma \right)
\mu }{12\left( 1-q\right) ^{2}\left( 1+2\gamma \right) ^{2}}.
\end{equation}

Evaluating $P_{1}^{1}\left( q,\mu \right) $ at the critical point
$X_{\gamma
}$, given by Eq. (\ref{critp}), we obtain %%%%%%%%%%
%Am schimbat semnele in formula care urmeaza
%%%%%%%%%
\begin{equation}
P_{1}^{1}\left( X_{\gamma }\right) = \frac{8\gamma^3 + 66\gamma -47} {%
4\left(2+\gamma \right)^2\left(1+2\gamma \right)}.
\end{equation}

\begin{figure}[!ht]
\centering
\includegraphics[height=5cm]{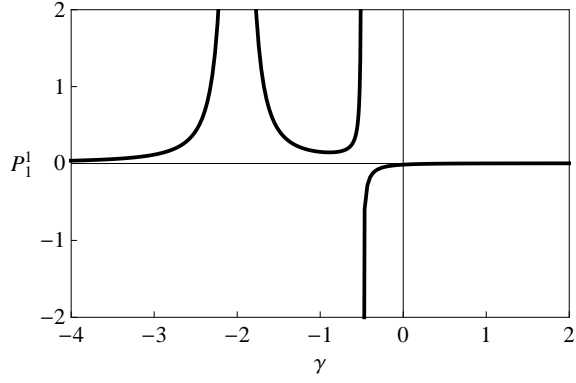}
\caption{The deviation curvature tensor $P_{1}^{1}\left(X_{\protect\gamma }\right) $ as a function of $\protect\gamma $.}
\label{figf3}
\end{figure}

The plot of the function $P_{1}^{1}\left( X_{\gamma }\right) $ is
represented in Fig.~\ref{figf3}.
%%%%%%%%%%%%%%%%%%%%%%%%%%%%%%%%%%%%%%%%
%%%%%%%%%%%%%%%%%%%%%%%%%%%%%%%%%%%%%%%%%

Using the discussion of Section~\ref{linstab}, our main results on the Linear stability and Jacobi stability of the critical point $X_{\gamma}$ of the vacuum field equations in the brane world models can be summarized in Table~\ref{table2}.

\begin{table}
\begin{tabular}{|c|ccccccc|}
\hline
$\gamma$ & $-\infty$\hfill\mbox{} & $-0.5\ $\mbox{} & & 0.67 & & 1 & \hfill$+\infty$\\
\hline
{$P_1^1(X_{\gamma})$} & $+$ & \big| & $-$ & \big| & $+$ & \big| & $+$ \\
\hline
{Linear stability } & Saddle & \big| & Stable & \big| & Stable & \big| &
Saddle \\
{of $X_{\gamma}$} & point & \big| & spiral & \big| & node & \big| & point \\
\hline
{Jacobi stability } & Jacobi & \big| & Jacobi & \big| & Jacobi & \big| &
Jacobi \\
{of $X_{\gamma}$} & unstable & \big| & stable & \big| & unstable & \big| & unstable \\
\hline
\end{tabular}
\caption{Linear and Jacobi stability of the stable point $X_{\gamma} $ of the vacuum field equations on the brane.}
\label{table2}
\end{table}

%%%
%%% DYNAMICAL DARK ENERGY
%%%

\section{Dynamical dark energy models}
\label{sec7}

Let us consider a dynamically evolving scalar field with exponential potential $V=V_{0}\exp(-\lambda\kappa\phi)$ to model dark energy in a spatially flat Friedmann-Robertson-Walker (FRW) universe~\cite{Co1997}. Moreover, let us assume that the universe also contains a dark matter component, modeled as a pressure-less perfect fluid, $p_\gamma = (\gamma-1)p_\gamma$, with $\gamma=1$. Henceforth we will denote the dark matter density as $\rho_{\mathrm{dm}}$. The Hubble function, describing the dynamics of the expansion of the Universe, is denoted by $H$.

The evolution equations of this cosmological model are
\begin{align}
  \dot{H} &= - \frac{\kappa^2}{2}(\rho_{\mathrm{dm}} + \dot{\phi}^{2}),
  \label{eomH} \\
  \dot{\rho}_{\mathrm{dm}} &= -3H\rho_{\mathrm{dm}},
  \label{eomrho} \\
  \ddot{\phi} &= -3H\dot{\phi}-\frac{dV}{d\phi},
  \label{eomphi}
\end{align}
subject to the Friedmann constraint
\begin{align}
  H^{2}=\frac{\kappa^2}{3}(\rho_{\mathrm{dm}}+\frac{1}{2}\dot{\phi}^2+V).
\end{align}
where $\kappa^2 = 8\pi G/c^4$.

Following~\cite{Co1997}, we introduce the dimensionless variables $u$ and $v$ defined by
\begin{equation*}
  u:=\frac{\kappa \dot{\phi}}{\sqrt{6}H},\qquad v:=\frac{\kappa \sqrt{V}}{%
\sqrt{3}H}.
\end{equation*}
Moreover, let $N=\ln (a)$, the evolution equations (\ref{eomH})--(\ref{eomphi}) can be written as a two-dimensional autonomous system of differential equations
\begin{align}
  \frac{du}{dN}& = -3u + \lambda \sqrt{\frac{3}{2}}v^{2} + \frac{3}{2} u[1+u^{2}-v^{2}],
  \label{dyn1} \\
  \frac{dv}{dN}& = -\lambda \sqrt{\frac{3}{2}}uv + \frac{3}{2}v[1+u^{2}-v^{2}],
  \label{dyn2}
\end{align}
while the Friedmann constraint now reads
\begin{equation*}
\frac{\kappa^{2}\rho_{\mathrm{dm}}}{3H^{2}}+u^{2}+v^{2}=1,
\end{equation*}
where the prime denotes differentiation with respect to $N$ as defined above.

The critical points of this dynamical system and the results of the linear stability analysis are given in Table~\ref{cpoints}.

\renewcommand{\baselinestretch}{1.75}
\begin{table}[!htb]
\centering
\begin{tabular}[t]{|l|c|c|c|c|}
\hline
Point & $u_*$ & $v_*$ & Existence & Stability \\ \hline\hline
A & $0$ & $0$ & $\forall\lambda$ & Saddle point \\ \hline
B$_{+}$ & $1$ & $0$ & $\forall\lambda$ & Unstable node $\lambda < \sqrt{6}$
\\
&  &  &  & Saddle point $\lambda > \sqrt{6}$ \\ \hline
B$_{-}$ & $-1$ & $0$ & $\forall\lambda$ & Unstable node $\lambda > -\sqrt{6}$
\\
&  &  &  & Saddle point $\lambda < -\sqrt{6}$ \\ \hline
C & $\frac{\sqrt{3/2}}{\lambda }$ & $\frac{\sqrt{3/2}}{%
\lambda}$ & $\lambda^2 > 3$ & Stable node $3 < \lambda^2 < 24/7$ \\
&  &  &  & Stable spiral $\lambda^2 > 24/7$ \\ \hline
D & $\frac{\lambda }{\sqrt{6}}$ & $\sqrt{1-\frac{\lambda ^2}{6}}$ & $%
\lambda^2<6$ & Stable node $\lambda^2 < 3$ \\
&  &  &  & Saddle point $3 < \lambda^2 < 6$ \\ \hline
\end{tabular}
\caption{Critical points and linear stability results for the dynamical dark energy models.}
\label{cpoints}
\end{table}
\renewcommand{\baselinestretch}{1}

\subsection{Lyapunov method}

For this particular system under investigation, a suitable Lyapunov function
is found to be
\begin{align}
  V(u,v)=(x-x_{\ast })^{2}+2(y-y_{\ast })^{2}
\end{align}
Clearly, $V$ has a global minimum near $(u_{\ast },v_{\ast })$ and we have
\begin{align}
  \frac{dV}{dN} = (\partial_{u}V)\frac{du}{dN} +(\partial_{v}V)\frac{dv}{dN} = 2(u-u_{\ast })\frac{du}{dN} + 4(y-y_{\ast })\frac{dv}{dN},
\end{align}
with $du/dN$ and $dv/dN$ from the autonomous system (\ref{dyn1})--(\ref{dyn2}). The points $A$ and $B_{\pm }$ are saddle points when analyzed using linear stability analysis and therefore we do not expect to find a Lyapunov function for these points. Every function we analyzed did not satisfy the required criteria. However, for the points C and D, the identified function allows us the establish nonlinear stability.

For the point C we analyze the eigenvalues of the Hessian of $dV/dN(u,v)$ at
that point. The two eigenvalues of $\mathrm{Hess}(dV/dN(u,v))$ are given by
\begin{align}
  \mu_{\pm} = -3 - \frac{9}{\lambda^2} \pm
  \frac{3}{\lambda^2}\sqrt{\lambda^4 - 18 \lambda^2 + 90}.
\end{align}
In the region where both eigenvalues are negative, the function $dV/dN$ has a maximum. Since $v>0$, this is equivalent with $\lambda^2 > 27/8$.

For the point D the two eigenvalues are given by
\begin{align}
  \mu_{\pm} = -18 + 4 \lambda^2 \pm
  \sqrt{36+6\lambda^2-\lambda^4}.
\end{align}
Both eigenvalues are negative if $\lambda^2 < 48/17$.

\subsection{Jacobi stability analysis}

In order to analyze the dynamical system (\ref{dyn1})--(\ref{dyn2}) using the Jacobi method, we first of all have to write this system as one second order differential equation of the form
\begin{align}
  x'' + 2 G^{1}(x,y) = 0,
  \label{eqfin}
\end{align}
where $x = u$, $y = u'$ and the prime denotes differentiation with respect to $N$.

Let us start by solving Eq.~(\ref{dyn1}) with respect to $v^{2}$ which gives
\begin{align}
  v^{2} = \frac{2u' + 3u - 3u^{3}}{\sqrt{6}\lambda-3u},
\end{align}
which allows us to compute $v'$ in terms of $u$ and $u'$. Now, taking the derivative of Eq.~(\ref{dyn1}) and substituting in the expressions of $y^{2}$ and $yy'$ gives the second order ordinary differential equation $F(u'',u',u)=0$. With $x = u$, $y = u'$ we find that $G^1(x,y)$ is given by
\begin{multline}
  G^{1}(x,y) = \frac{3}{4(\sqrt{6}\lambda-3u)}\Bigl[
    6 (3 x^2 - 3 x^4 + 3 x y + y^2) \\+
    \lambda \sqrt{6} (-3 x - 3 x^3 + 6 x^5 - y - 7 x^2 y) +
    \lambda^2 (6 x^2 - 6 x^4 + 4 x y)\Bigr].
\end{multline}

As a first step in the KCC stability analysis of the dark energy model described by Eq.~(\ref{eqfin}) we obtain the nonlinear connection $N_{1}^{1}=\partial G^{1}/\partial y$ which is given by
\begin{align}
  N_{1}^{1} = \frac{3}{4(\sqrt{6}\lambda-3u)}\Bigl[
    6 (3 x + 2 y) - \lambda \sqrt{6} (1 + 7 x^2) + 4 \lambda^2 x\Bigr].
\end{align}
The associated Berwald connection $G_{11}^{1} =  \partial N_{1}^{1}/\partial y$ can be obtained as
\begin{align}
  G_{11}^{1} = \frac{9}{\sqrt{6}\lambda-3u}.
\end{align}

Finally, the second KCC invariant, or the deviation curvature tensor $P_{1}^{1}$, defined as
\begin{align}
  P_{1}^{1} = -2\frac{\partial G^{1}}{\partial x} - 2G^{1}G_{11}^{1} +
  y\frac{\partial N_{1}^{1}}{\partial x} + N_{1}^{1}N_{1}^{1},
\end{align}
and is found to be
\begin{align}
  P_{1}^{1}(x,y) = \frac{3}{8(\sqrt{6}\lambda-3u)^2}\Bigl[
  54x^2 + \lambda a_1(x,y) + \lambda^2 a_2(x,y) +
  \lambda^3 a_3(x,y) + \lambda^4 a_4(x,y) \Bigr].
\end{align}
where the four functions $a_i(x,y)$ are given by
\begin{align}
  a_1(x,y) &= 6 \sqrt{6} (-15 x - 9 x^3 + 12 x^5 - 5 y - 7 x^2 y),\\
  a_2(x,y) &= 81 + 414 x^2 - 279 x^4 + 168 x y,\\
  a_3(x,y) &= 4 \sqrt{6} (-15 x + 3 x^3 - 2 y),\\
  a_4(x,y) &= 24 x^2.
\end{align}

To understand the stability at the critical points A--D, we evaluate $P_{1}^{1}$ and obtain

\begin{align}
  P_1^1(\mathrm{A}) &=
  \Bigl(\frac{3}{2}\Bigr)^4 > 0,\\
  P_1^1(\mathrm{B}_{\pm}) &=
  \frac{3}{2} (\lambda \mp \sqrt{3/2})^2 \geq 0
  \qquad \forall\, \lambda,\\
  P_1^1(\mathrm{C}) &=
  \frac{9}{2}\Bigl(-\frac{7}{8}+\frac{3}{\lambda^2}\Bigr),\\
  P_1^1(\mathrm{D}) &=
  \Bigl(\frac{\lambda}{2}\Bigr)^4 > 0\qquad \forall\, \lambda \neq 0.
\end{align}

It is immediately clear that the only point that can be Jacobi stable is in fact point C. If $\lambda^2 > 24/7$ we find that $P_1^1(C) < 0$ and hence in that case the point is stable. Our results for the uncoupled models are summarized in Table~\ref{table_cosm}.

\renewcommand{\baselinestretch}{1.75}
\begin{table}[!ht]
\begin{tabular}{|c|c|c|c|c|}
\hline
Point & $\lambda$ & Linear & Jacobi & Lyapunov \\ \hline\hline
C & $3 < \lambda^2 < 27/8$ & Stable node & unstable & inconclusive \\
& $27/8 < \lambda^2 < 24/7$ & Stable node & unstable & asymptotically stable \\
& $\lambda^2 > 24/7$ & Stable spiral & stable & asymptotically stable \\ \hline
D & $\lambda^2 < 48/17$ & Stable node & unstable & asymptotically stable \\
& $48/17 < \lambda^2 < 3$ & Stable node & unstable & inconclusive \\
& $3 < \lambda^2 < 6$ & Saddle point & unstable & inconclusive \\ \hline\hline
\end{tabular}
\caption{Stability analysis of the dynamic dark energy models.}
\label{table_cosm}
\end{table}
\renewcommand{\baselinestretch}{1}

\section{Discussions and final remarks}
\label{sec8}

In the present paper we reviewed two basic stability analysis methods -- (Lyapunov's) linear stability analysis and the Jacobi stability analysis, respectively. We considered the stability properties of several dynamical systems that play important roles in gravitation and cosmology -- Newtonian polytropes, described by the Lane-Emden equation, general relativistic fluid spheres, the vacuum gravitational field equations of brane world models, and dynamical dark energy models, respectively. For all cases we used both methods to investigate stability, (Lyapunov) linear stability analysis and Jacobi stability analysis, or the KCC theory.  The study of stability has been done by analyzing the behavior of steady states of the respective dynamical system. Linear stability analysis involves the linearization of the dynamical system via the Jacobian matrix of a non-linear system, while the KCC theory addresses stability of a whole trajectory in a tubular region \cite{Sa05}.

By using the KCC theory we have been able to derive some physically relevant results for each of the considered systems. In the case of the Newtonian polytropes from the Jacobi stability condition we obtain a restriction of the ratio of the internal energy of the star and of its gravitational energy in terms of ratio of the density and of the mean density of the star. In the case of the general relativistic fluid sphere the Jacobi stability condition gives a physical range of the mass-radius ratio for general relativistic compact stellar type objects. In the case of the brane world models we have shown that the vacuum on the brane is Jacobi unstable for most of the values of the parameter $\gamma $, with the stability region reduced to a very narrow range of $\gamma $. For all other values of $\gamma $, the vacuum on the brane is unstable, in the sense that the trajectories of the structure equations will disperse when approaching the origin of the coordinate system. The Jacobi stability analysis also imposes some strong restrictions on the parameter $\lambda $ describing the properties of dark energy in the dynamic dark energy models. 

It was shown in \cite{Sa05} that the Jacobi stability of a dynamical system can be regarded as the {\it robustness} of the system to small perturbations of the whole trajectory. Thus, Jacobi stability is a very conveyable way of regarding the resistance of limit cycles to small perturbation of trajectories. On the other hand, we may regard the Jacobi stability for other types of dynamical systems (like the ones considered in the present paper) as the resistance of a whole trajectory to the onset of chaos due to small perturbations of the whole trajectory. This interpretation is based on the generally accepted definition of chaos, namely a compact manifold $M$ on which the geodesic trajectories deviate exponentially fast. This is obviously related to the curvature of the base manifold (see Section \ref{3}). The Jacobi (in)stability is a natural generalization of the (in)stability of the geodesic flow on a differentiable manifold endowed with a metric (Riemannian or Finslerian) to the non-metric setting. In other words, we may say that Jacobi unstable trajectories of a dynamical system behave chaotically in the sense that after a finite interval of time it would be impossible to distinguish the trajectories that were very near each other at an initial moment.

We have also found that for all the dynamical systems considered in the present paper there is a good correlation between the linear stability of the critical points, and the robustness of the corresponding trajectory to a small perturbation. Indeed, in the case of the gravitational field equations of the vacuum on the brane, for small values of the parameter $\gamma$ the saddle point is also Jacobi unstable ($\gamma<-0.5$ and $\gamma>1$), while the stable spiral obtained for $-0.5<\gamma<0.674865$ is also robust to small perturbations of all trajectories. It is also interesting to remark for the same system that for the interval $0.674865<\gamma<1$, the stable node is actually Jacobi unstable. In other words, even though the system trajectories are attracted by the critical point $X_{\gamma}$ one has to be aware of the fact that they are not stable to small perturbation of the whole trajectory. This means that one might witness chaotic behavior of the system trajectories before they enter a neighborhood of $X_{\gamma}$. We have here a sort of stability artefact that cannot be found without using the powerful method of Jacobi stability analysis. In the present review we have presented in detail the main theoretical tools necessary for an in depth analysis of the stability properties of dynamical systems that play a fundamental role in our understanding of nature. 

\section*{Acknowledgments}
The work of TH was supported by the GRF grant number 701808P of the government of the Hong Kong SAR.

\end{document}